\documentclass[pra,showpacs,twocolumn,superscriptaddress]{revtex4-2}

\usepackage{amsmath}
\usepackage{amssymb}
\usepackage{caption}
\usepackage{graphicx}   % need for figures
\usepackage{color}      % use if color is used in text
\usepackage{hyperref}   % use for hypertext links, including those to external documents and URLs
\usepackage{dcolumn}

\begin{document}
\title{State Transfer and Entanglement between Two- and Four-Level Atoms in A Cavity}

\author{Si-Wu Li}
\affiliation{School of Physics, Sun Yat-Sen University, Guangzhou 510275, China}
\author{Tianfeng Feng}
\affiliation{School of Physics, Sun Yat-Sen University, Guangzhou 510275, China}
\author{Xiao-Long Hu}
\affiliation{School of Physics, Sun Yat-Sen University, Guangzhou 510275, China}
\author{Ze-Liang Xiang}
\email{xiangzliang@mail.sysu.edu.cn}
\affiliation{School of Physics, Sun Yat-Sen University, Guangzhou 510275, China}
\author{Xiaoqi Zhou}
\email{zhouxq8@mail.sysu.edu.cn}
\affiliation{School of Physics, Sun Yat-Sen University, Guangzhou 510275, China}

\date{\today}
\begin{abstract}
Qudits with a large Hilbert space to host quantum information are widely utilized in various applications, such as quantum simulation and quantum computation, but the manipulation and scalability of qudits still face challenges. Here, we propose a scheme to directly and locally transfer quantum information from multiple atomic qubits to a single qudit and vice versa in an optical cavity. With the qubit-qudit interaction, our scheme can transfer quantum states efficiently and measurement-independently. In addition, this scheme can be extended to the non-local case, where a high-dimensional maximal entangled state with asymmetric particle numbers can be robustly generated for realizing long-distance quantum communication. Such an information interface for qubits and qudit may have enlightening significance for future research on quantum systems in hybrid dimensions.
\end{abstract}
\maketitle
%================================================================================
%================================================================================

\section{Introduction}
At present, most theories and experiments about quantum information are based on two-level systems, so-called qubits. However, in the microscopic world, atoms~\cite{parallelism_atom_qudit,quantum_search_atom_qudit,entanglement_squeezing_D-level,qutrit_simulation,qudit_ion}, photons~\cite{photonic_molecules_vibration,qudit-based_processer_photon,qudit_photons_quantum-walks,qudit-based_universal_computation}, and solid-state systems~\cite{Emulation_Superconducting_qudit,Scrambling_superconducting_qutrit,spin-qudit_silicon_carbide,spin-qudit_molecular} actually possess more than two eigenstates, which allows us to store and handle $d$-dimensional quantum information in a single `qubit' named qudits. Compared to qubits, such systems have exponentially increased information capacity and show enormous potential in constructing high-dimensional quantum gates to boost the performance of quantum computation~\cite{optimal_d-level,simplify_higher-dimensional,qudit_toffoli}. Qudits can also locally implement complex Hamiltonian in quantum simulations, such as the discovery of new gauge fields for new physics~\cite{SO(3)_simulation,qudit_simulation_rydberg}, and quantum chemical dynamics involving molecular or atomic vibrations~\cite{MQB_chemical_dynamics}. Furthermore, one can use qudit systems to realize entanglement in more complex structures~\cite{multilevel_entanglement,intergrated_optics_entangled,OAM_entangled,high_entangled_chip,GHZ_transmon,superconducting_qutrit} and stronger Bell non-locality~\cite{qudit_bell}, and to explore the boundaries of entanglement of physical systems.
 
In order to manipulate qudits, coupling well-studied qubits to qudits is a promising way. Thus, it is necessary to study the connection and also state transfer between the particles with different dimensions, such as multiple qubits and qudits, in a hybrid-dimensional system~\cite{212}. %One can encode qubits' information to a single qudit or vice versa, i.e., transfer quantum state between them~\cite{212}. 
Recently, various approaches were proposed to implement state transfer in the same-dimensional system by employing the cavity-assisted Raman process~\cite{PRL_state_transfer,remote-state-transfer_microwave-photons} and resonant cavities~\cite{state_transfer_resonant_cavity,qudit_transfer_transmon}, which all require cavities with high quality factors. Besides, the direct interaction between qubits can also be used to realize state transfer without the restriction of the cavity's quality~\cite{transfer_QDs, CNOT_SWAP_spins}, and it is widely adopted in the solid-state quantum system for fast and scalable quantum information processing. Inspired by these schemes, constructing the interaction between qubits and qudits in quantum electrodynamics (QED)~\cite{superconducting_cQED,RMP_cirQED,ion_simulator,ultracold_QED,kiraz2001cavity,vuvckovic2003photonic}, which describes photon-atom or atom-atom interaction and has been used to realize quantum logic operations~\cite{simple_cQED_2bit,QD_cQED_gate,SQID_gate},  can also achieve the state transfer. Moreover, long-distance quantum teleportations~\cite{PRL_tele,Nature_DLCZ,qudit_teleportation_linear-optics,STRAP_qudit,qudit_teleportation_photonic,teleportation_ququart,qudit_teleportation_GHZ} are based on quantum entanglement channels, and those entangled states are also prepared in same-dimensional quantum objects. Researchers have proposed a number of schemes to prepare entanglements between atoms in QED systems~\cite{two-atom_cQED,multiparticle_cQED,W_cQED}. To realize a long-distance communication channel, we can also utilize QED to generate a special entanglement in a hybrid-dimensional quantum system.
 
In this paper, we propose a cavity QED-based scheme to construct photon-induced qubit-ququart interaction that enables us to implement quantum state transfer between two qubits and a single ququart. Using this coherent interaction and photon-number-dependent shift, arbitrary states in two qubits can be perfectly encoded into the ququart and vice versa. Our proposed scheme can also be used to robustly and noise-resistantly generate a maximum entanglement between two-dimensional and four-dimensional systems, i.e., to prepare an asymmetric maximally entangled state (the AMES or (2, 2, 4) entangled state)~\cite{224_photon}. Such an entangled state can be used to establish a long-distance information interface for atoms possessing different levels. Unlike previous implementations on linear optics, our approach can prepare the (2, 2, 4) entangled state efficiently and determinately. We believe our scheme can be extended to the case of multiple qubits and qudits of more dimensions as the structural complexity of the light fields increases.

The layout of this paper is as follows. In Sec.~\ref{SecII}, we introduce the system of the proposed scheme and derive an effective Hamiltonian for transferring the information between hybrid-dimensional systems. Then we discuss the state-transfer process in two different cases with or without the AC-Stark phase shift. Sec~\ref{SecIII} further studies how to utilize a modified system to prepare AMES in high fidelity and show numerical results about how errors of parameters affect the fidelity of the state's preparation. In Sec.~\ref{SecIV}, we introduce a long-distance quantum teleportation protocol as an application of the atomic AMES. Finally, we make a conclusion for our state transfer and entanglement preparation scheme, and a vision of the future.
%================================================================================
%================================================================================

\section{Direct Information Transfer Between Two Qubits and a Ququart}\label{SecII}

Our target is to construct a cavity QED system, in which quantum information can be transferred from two qubits to a ququart, i.e.,
\begin{equation}\label{eq_transfer}
\begin{split}
	|\psi_{0,1}\rangle =&~(C_{gg}|gg\rangle+C_{eg}|eg\rangle+C_{ge}|ge\rangle+C_{ee}|ee\rangle) \otimes |1\rangle \\
	\rightarrow|\psi_{\rm tar}\rangle =&~|gg\rangle \otimes (C_{gg}|1\rangle+C_{ge}|2\rangle+C_{eg}|3\rangle+C_{ee}|4\rangle),
\end{split}
\end{equation}
where the first and second entries in the kets represent states of qubits A and B, respectively, while the third entry represents the states of the ququart.

\begin{figure}
	\centering
	\includegraphics[width=\linewidth]{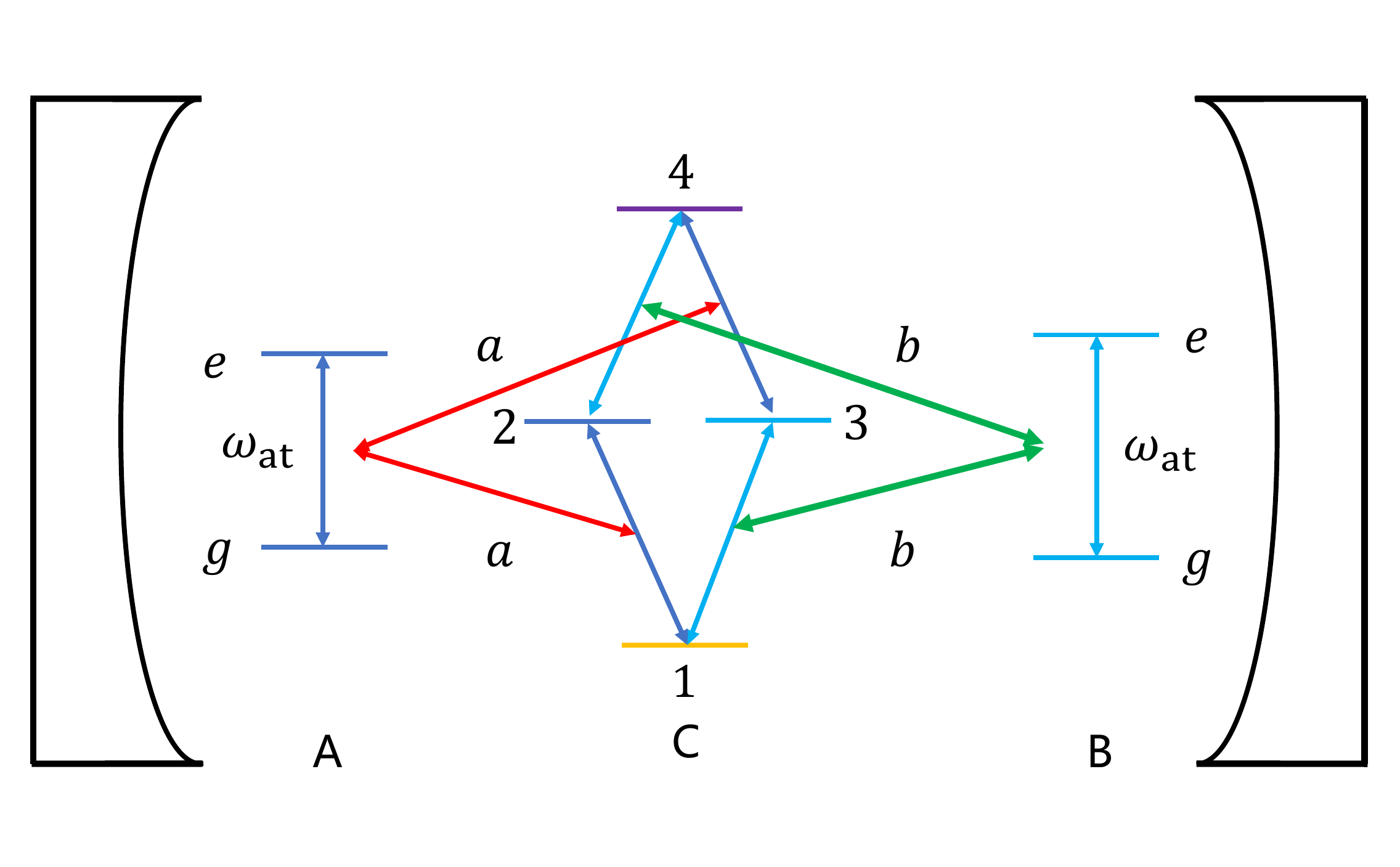}
	\captionsetup{justification=raggedright}
	\caption{Sketch of the atomic level structure and coupling field mode in our proposed scheme. Two-level atoms A and B interact with polarized field modes $a$ (red arrow) and $b$ (green arrow) respectively. The four-level atom contains orthogonal transition dipoles interacting with polarized field modes.}
	\label{fig.1}
\end{figure}

 We consider a system consisting of two qubits and one ququart coupled with a same cavity via two orthogonal polarized photonic modes $a$ and $b$, whose exciting frequencies are $\omega_{\rm op}$. The atomic qubits have transition dipoles with degenerate excited frequencies $\omega_{\rm at}$, and couple to modes $a$ and $b$, respectively. The single ququart atom possesses two single-exciton states and one biexciton state, whose transitions also interact with two cavity modes. Figure~\ref{fig.1} depicts the level structures of qubits and the ququart. The system's Hamiltonian can be written as
\begin{equation}\label{eq_H_total}
\begin{split}
	H=&~H_0+H_I,\\
	H_0=&~\omega_{\rm ph} a^{\dagger}{a}+\omega_{\rm at}(\sigma_{ A}^{\dagger}\sigma_{ A}+|2\rangle \langle2|+|4\rangle\langle4|)\\
	&~+\omega_{\rm ph} b^{\dagger}b+\omega_{\rm at}(\sigma_{ B}^{\dagger}\sigma_{ B}+|3\rangle\langle 3|+|4\rangle\langle4|),\\
	H_{\rm I}=&~g_{ A}(a^{\dagger}D_{ A}+D_{ A}^{\dagger}a)+g_{ B}(b^{\dagger}D_{ B}+D_{ B}^{\dagger}b).\\
\end{split}
\end{equation}
Here, we consider a Jaynes-Cummings type interaction between multiple atoms and cavity modes with coupling strengths $g_A$ and $g_B$, respectively, which are identical for the same polarization. The collective dipole operators read
\begin{equation}
\begin{split}
		D_{ A}=\sigma_{ A}+|1\rangle \langle2|+|3\rangle\langle4|,\\
		D_{ B}=\sigma_{ B}+|1\rangle \langle3|+|2\rangle\langle4|.\\
\end{split}
\end{equation}
where $\sigma_n=|g\rangle_n \langle e|_n$, which is the lower operator of the two-level system. 
%================================================================================
%================================================================================

 When the energy of the cavity fields does not match the exciting energy of the atomic system, the effective interaction would be weakened, and there is no energy exchange between them. Cavity modes can be virtually excited, and the second-order interaction dominates the system's dynamics, where qubits and the ququart exchange their excitations. In this case, we can use this cavity QED model to implement the quantum state transfer, while the decoherence of the cavity is suppressed. We perform the Schrieffer-Wolf transformation~\cite{SWtransformation},
\begin{equation}
	 U={\rm exp}[-\frac{g_{ A}}{\Delta}(D_{ A}^{\dagger}a-a^{\dagger}D_{ A})-\frac{g_{ B}}{\Delta}(D_{ B}^{\dagger}b-b^{\dagger}D_{ B})],
\end{equation}
on $H$ to adiabatically eliminate cavity modes in the system's evolution, where the detuning satisfies $\Delta=\omega_{\rm ph}-\omega_{\rm at}\gg g_A, ~g_B$. After the transformation, the effective Hamiltonian of the system becomes
\begin{equation}\label{eq_H_eff}
\begin{split}
	H_{\rm eff}=H_0-\frac{g_{ A}^2}{\Delta}[D_{ A}^{\dagger}D_{ A} +D_{ A}^{ z} a^{\dagger}a]-\frac{g_{ B}^2}{\Delta}[D_{ B}^{\dagger}D_{ B} +D_{ B}^{ z} b^{\dagger}b],
\end{split}
\end{equation}
where
\begin{equation}
\begin{split}
	D_{ A}^{ z}=&~\sigma_{ A}^{\dagger}\sigma_{ A}-\sigma_{ A}\sigma_{ A}^{\dagger}+|4\rangle\langle4|-|3\rangle\langle3|+|2\rangle\langle2|-|1\rangle\langle1|,\\
	D_{ B}^{ z}=&~\sigma_{ B}^{\dagger}\sigma_{ B}-\sigma_{ B}\sigma_{ B}^{\dagger}+|4\rangle\langle4|-|2\rangle\langle2|+|3\rangle\langle3|-|1\rangle\langle1|.
\end{split}
\end{equation}
The second and fourth terms in Eq.~\ref{eq_H_eff} describe the qubit-ququart interaction, while the third and fifth terms denote the photon-number-dependent shift. 
 
With the effective coupling strength, $\lambda \equiv -g_A^2/\Delta=-g_B^2/\Delta$ and the photon-number-dependent phase shift in $H_{\rm eff}$, this cavity QED system can achieve the state transfer. First, we assume that $a$ and $b$ modes initially are both in vacuum states so that the effective Hamiltonian in Eq.~\eqref{eq_H_eff} can be projected into the photon-vacuum subspace, i.e.,
\begin{equation}\label{eq_H_eff_vac}
	H_{\rm  eff,vac}=H_0 +\lambda(D_{ A}^{\dagger}D_{ A}+D_{ B}^{\dagger}D_{ B}).
\end{equation}
Because Eq.~\eqref{eq_transfer} contains states in different subspaces with the particle-exchange interaction, the state evolution could be written as
\begin{equation}
\begin{split}
	|\psi(t)\rangle =&~C_{gg}|gg1\rangle + \\
	&~C_{eg} [\alpha_{eg}(t) |eg1\rangle+\beta_{eg}(t) |gg2\rangle]+\\
	&~C_{ge} [\alpha_{ge}(t) |ge1\rangle+\beta_{ge}(t) |gg3\rangle]+\\
	&~C_{ee} [\alpha_{ee}(t) |ee1\rangle+\beta_{ee}(t) |ge2\rangle+\\
	&~\theta_{ee}(t)|eg3\rangle + \mu_{ee}(t) |gg4\rangle],\\
\end{split}
\end{equation}
where, $\alpha_{\mu \nu}$ and $\beta_{\mu \nu}$ represent normalized amplitudes for their corresponding subspaces. Note that we have omitted the cavity's degree of freedom. With $H_{\rm eff,vac}$, the two-qubit state will be transferred to the ququart at $t=\pi/2\lambda$,
\begin{equation}\label{eq_H_eff_vac_transfer}
\begin{split}
	|\psi(\frac{\pi}{2\lambda})\rangle=&~{\rm exp}(-iH_{\rm eff,vac}\frac{\pi}{2\lambda})|\psi_{0,1}\rangle\\
    =&~|gg\rangle \otimes (C_{gg}|1\rangle-C_{eg}|2\rangle-C_{ge}|3\rangle+C_{ee}|4\rangle).
\end{split}
\end{equation}
It indicates additional minus signs in the final state, which will noticeably decrease the transfer fidelity (see Appendix~\ref{AppA}). For correcting these minus signs, single-qubit phase operations on qubits A and B should be performed, respectively, before the cavity-induced interaction. Then the cavity QED model without the photonic excitation implements the local state-transfer operation with 100\% fidelity.

Furthermore, using the photon-number-dependent phase shift, the system can also achieve a perfect information transfer without single-qubit operations. Initially, the double-mode single-photon state is prepared in the cavity, and $H_{\rm eff}$ in the single-photon subspace is given by
\begin{equation}\label{eq_H_eff_ph}
	H_{\rm  eff,ph}=H_0 +\lambda(D_{ A}^{\dagger}D_{ A}+D_{ B}^{\dagger}D_{ B}+D_{A}^{z}+D_{B}^{z}).
\end{equation}
The final state of the evolution is
\begin{equation}\label{eq_transfer_4to4}
\begin{split}
	|\psi'({\frac{\pi}{2\lambda}})\rangle = &~{\rm exp}(-iH_{\rm eff,ph}\frac{\pi}{2\lambda})|\psi_{0,1}\rangle\\
    =&~|gg\rangle \otimes (C_{gg}|1\rangle+C_{eg}|2\rangle+C_{ge}|3\rangle+C_{ee}|4\rangle) .
\end{split}
\end{equation}
Finally, a perfect information transfer is realized without performing local qubit operations before the atomic interaction, however, the cavity's initial state should be prepared into a specific state.
%================================================================================
\begin{figure}
	\centering
	\includegraphics[width=\linewidth]{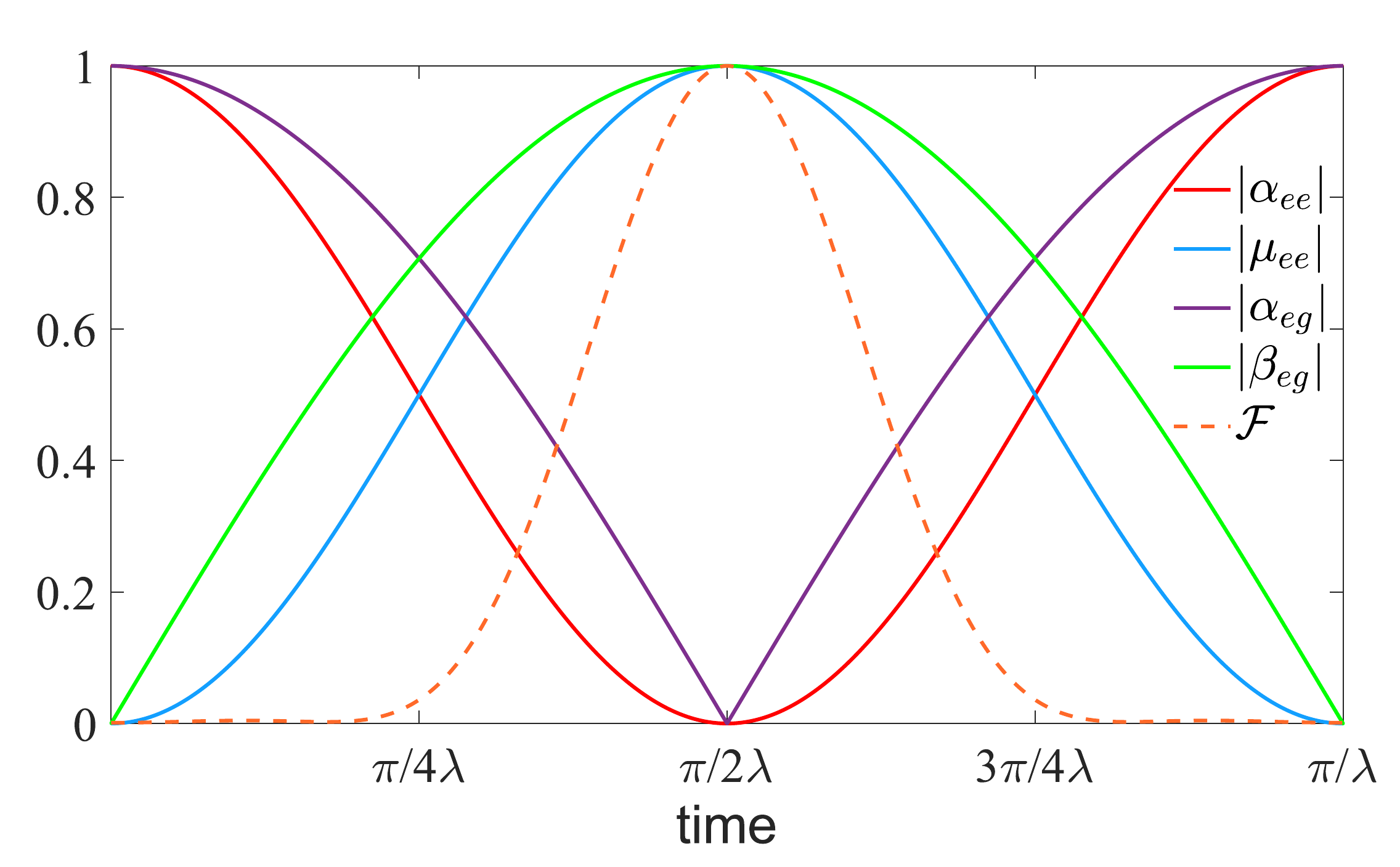}
	\captionsetup{justification=raggedright}
	\caption{Time-dependent Rabi-type oscillation of the state transfer between 2 qubits and a ququart, with the system's Hamiltonian $H_{\rm eff,ph}$. The transfer time is identical of single- and double-excitation subspaces, and amplitudes (solid lines) are transferred synchronously. The fidelity (orange dashed line), $\mathcal{F}$, indicates the phase difference between the transferred state and the ideal state.}
	\label{fig.2}
\end{figure}
%================================================================================

Figure~\ref{fig.2} reveals the states' evolution with the Hamiltonian $H_{\rm eff,ph}$, which contains the single-photon phase shift. At the initial time, $\alpha_{ge}=\alpha_{eg}=\alpha_{ee}=1$, and the two-qubit system is separated from the ququart. After a period of time, two systems are entangled and then are separated again at $t=\pi/2\lambda$, while the amplitude information, initially encoded in the two-qubit system, has been transferred to the single-ququart system. The time-dependent fidelity of the state transfer process defined as $\langle\psi(t)||\psi(T)\rangle|^2$ is also shown in Fig.~\ref{fig.2}. The fidelity reaches 100\% when $t=\pi/2\lambda$ along the system's evolution. The amplitude information is transferred back to the qubits after $t=\pi/2\lambda$, and it means that the inverse state transfer process can also be achieved with this approach.

In a ``2 to 1" state transfer process, one can implement the information interface for a hybrid-dimensional system to transfer two-body information on a single particle. In a reverse process, one can prepare arbitrary whatever entanglement or not two-body quantum state after preparing a single particle state without direct interaction between low-dimensional particles \cite{212}. 
 
This scheme can be expanded for a higher-dimensional system. A collective operator containing a two-level operator and multi-level ladder operators is given by
\begin{equation}
	J=\sigma_{ge}+ \Sigma_{\rho,\eta} \sigma_{\rho,\eta},
\end{equation}
where $\sigma_{\rho,\eta}=|\rho\rangle \langle \eta|$. The collective dipole will interact with electromagnetic modes to transfer the quantum information coherently, where these field modes are orthogonal, e.g., they can be photons carrying orbital angular momentum. 

%================================================================================
\begin{figure*}
	\centering
	\includegraphics[width=\linewidth]{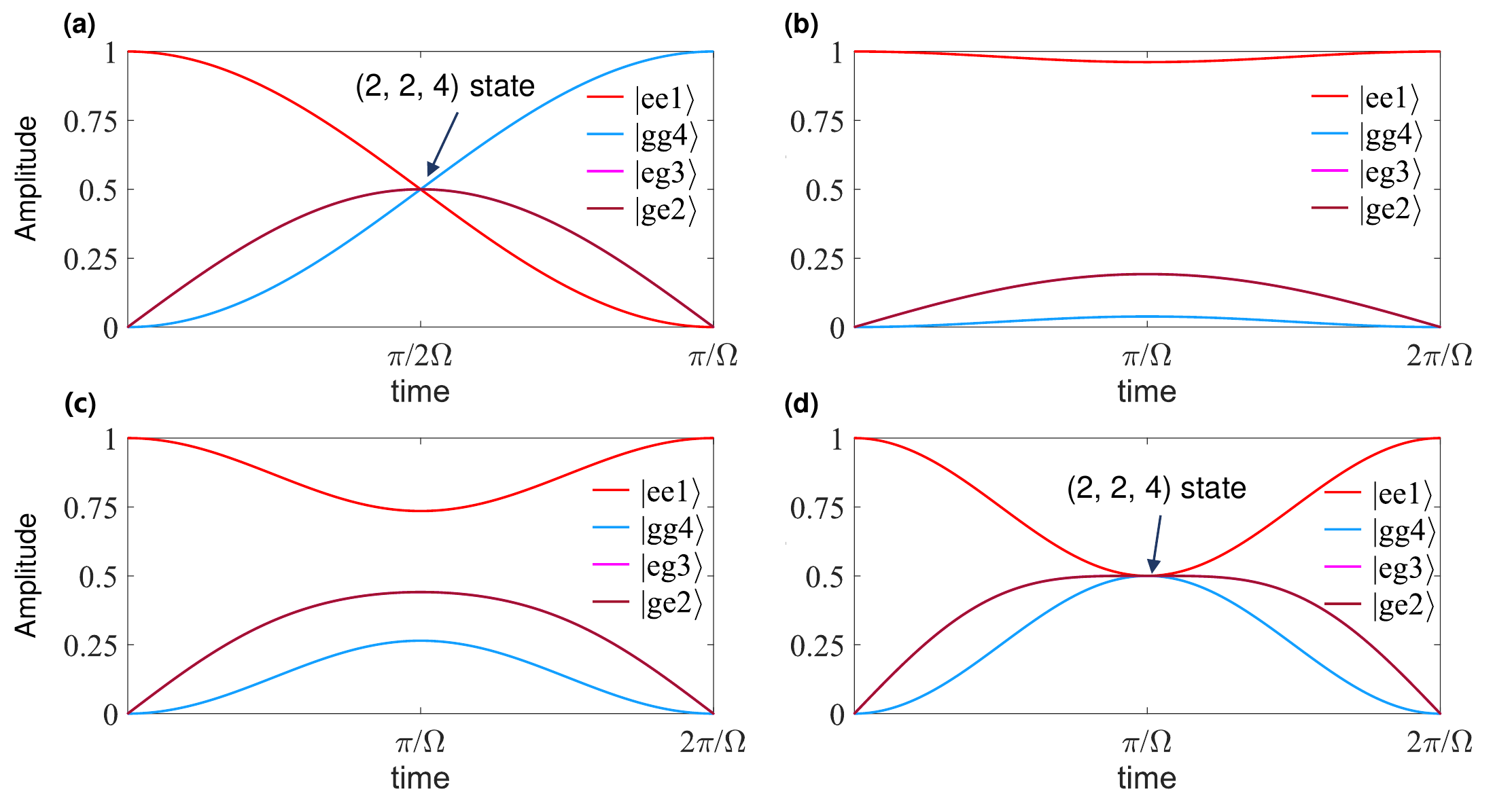}
	\captionsetup{justification=raggedright}
	\caption{Time-dependent evolution of the (2, 2, 4) entangled-state preparation in the cavity QED model. (a) shows the interaction process without the frequency mismatch. (b)-(d) display atomic interaction processes with the frequency mismatch as the coupling strength grows, i.e., $|\lambda'/\delta|=0.1, 0.3, 0.5$, respectively, and $\Omega=\sqrt{\delta^2+4\lambda'^2}$. The (2, 2, 4) state obtains as the amplitudes reach their extreme points, where the evolution is smooth and robust to the time error. Here, the amplitude of $|eg3\rangle$ (pink line) overlaps with $|ge2\rangle$ (purple line).}
	\label{fig.3}
\end{figure*}
%================================================================================

In experiments, the cavity with two orthogonal polarized modes can be realized by a circular resonator~\cite{chip_chiral-interface,isolator_cold-atoms,microring-resonators_linearly-polarized,optical_circulator,Exciton-Photon_strong_coupling_microcavity}, and the cavity photons are circularly polarized with $\pm1$ pseudo-spin. The ququart can be manufactured by artificial quantum-dots (QDs)~\cite{rmp_interfacing_qd,QD_biexciton,QD_biexciton_blockade}. The photon-atom coupling strength, $g=2\pi \times 15.2~{\rm Ghz}$, can be achieved in the quantum-dot cavity QED platform~\cite{silicon_microring}. When $\Delta=100g$, the time of the transfer will be $1.6\times 10^{-9}~{\rm s}$ and it means the cavity QED scheme realizes the ``2 to 1" state transfer much more efficiently than the linear optical implementation, which needs measurement and post selection to perform many-body quantum operations~\cite{212,224_photon}. The requirement on the quality factor is relaxed, because the qubit-ququart interaction is realized by the virtual-photon exchange. In the vacuum cavity case, the probability of exciting cavity modes is approximately $10^{-4}$, the quality factor $Q=3.9\times 10^{4}$ and $\omega_{\rm op}=2\pi\times 192~{\rm Thz}$~\cite{chip_chiral-interface}, therefore, the effective lifetime $\tau_{\rm cav}=Q/(10^{-4}\omega_{\rm op})=3.2 \times 10^{-7}~{\rm s}$, which is much longer than the state-transfer time. Electromagnetic fields with different excitation frequencies can also implement orthogonal modes, and the four-level system can also be realized by superconducting circuits~\cite{4-level_SQUID,4-level_SQUID-SR,EIT_4-level_SQC}. Our state-transfer scheme is accomplished by a local interaction system within a cavity, and we will expand our state-transfer scheme to generate a special entanglement state and the remote communication, which will be discussed in Sec.~\ref{SecIII} and Sec.\ref{SecIV}. 
%================================================================================
%================================================================================

\section{The Asymmetric Maximal Entanglement Preparation via Near-Resonant Interaction}\label{SecIII}

In this section, we discuss how to generate the maximal entanglement between two qubits and a ququart by exploiting the original and the modified system Hamiltonian. 

To increase the scale of entangled states for more complex applications, one can increase both the number and dimensions of particles, such types of entangled states have been demonstrated~\cite{malik2016multi,hu2020442}. Recently, an asymmetric maximal entangled state (AMES or (2, 2, 4) entangled state), consisting of two two-level systems and a single four-level system, i.e.,
\begin{equation}\label{eq_AMES}
	\frac{1}{2}(|001\rangle+|012\rangle+|103\rangle+|114\rangle),
\end{equation}
 has been probabilistically prepared in the photonic system~\cite{224_photon}. Because such an entangled state contains particles with different dimensions, it can be used as an interface for the information flow between quantum systems with different dimensions via the quantum state teleportation protocol.

Using our far-detuning cavity QED Hamiltonian in Eq.~\eqref{eq_H_eff_vac}, we can prepare the atomic AMES like Eq.~\eqref{eq_AMES} after an evolution time, $t=\pi/4\lambda$, and atomic entangled state is approximately pure because the cavity is virtually excited. However, the numerical result shown in Fig.~\ref{fig.3} (a) implies that the state evolution is not stable facing the error of the temporal control when the atomic system is in the AMES. 

Based on the system in Eq.~\eqref{eq_H_total}, we introduce a frequency mismatch $\delta$ between qubits and the ququart's exciton states, and the system Hamiltonian reads
\begin{equation}\label{eq10}
\begin{split}
	H_{\rm mis}=&~H_{\rm 0,mis}+H_{\rm I},\\
	H_{\rm 0,mis}=&~\omega_{\rm A}(\sigma_{ A}^{\dagger}\sigma_{ A}+|3\rangle \langle 3|+|4\rangle \langle 4|)+\omega_{\rm op} a^{\dagger}a \\
	&~+\omega_{ B}(\sigma_{ B}^{\dagger}\sigma_{ B}+|2\rangle \langle 2|+|4\rangle \langle 4|)+\omega_{\rm op} b^{\dagger}b,\\
	H_{\rm I}=&~g_{ A}(a^{\dagger}D_{ A}+D_{ A}^{\dagger}a)+g_{ B}(b^{\dagger}D_{ B}+D_{ B}^{\dagger}b),\\
\end{split}
\end{equation}
where $\omega_A=\omega_{\rm at}-\delta/2$, $\omega_B=\omega_{\rm at}+\delta/2$, and $\delta \ll \Delta$. Again, we apply the Schrieffer-Wolf transformation $e^{S_{\rm mis}}$ to $H_{\rm mis}$ and obtain an effective interaction between atoms, where 
\begin{equation}
\begin{split}
    S_{\rm mis}=&~\frac{g_A}{\Delta}(a^{\dagger}\sigma_A-\sigma_A^{\dagger}a)+\frac{g_B}{\Delta}(b^{\dagger}\sigma_B-\sigma_B^{\dagger}b)\\
	&~+\frac{g_A}{\Delta-\delta/2}[a^{\dagger}(|1\rangle \langle 2|+|3\rangle \langle 4|)-(|2\rangle \langle1|+|4\rangle \langle 3|)a]\\
	&~+\frac{g_B}{\Delta+\delta/2}[b^{\dagger}(|1\rangle \langle 3|+|2\rangle \langle 4|)-(|3\rangle \langle1|+|4\rangle \langle 2|)b].
\end{split}
\end{equation}
Then the final effective Hamiltonian changes to
\begin{equation}
	H_{\rm eff,mis}\approx H_{\rm 0,mis} 
	-\frac{g_A^2}{\Delta-\delta/2}D_A^{\dagger}D_A-\frac{g_B^2}{\Delta+\delta/2}D_B^{\dagger}D_B,
\end{equation}
where we have assumed that field modes are vacuum at the initial time. In the absence of frequency mismatch, the (2, 2, 4) state is always able to be generated at a certain time, no matter how the coupling strength is. When the frequency mismatch is introduced, the maximal entanglement can be prepared only as the atomic coupling is strong enough. In order to achieve the maximal entanglement, the effective coupling strength, $\lambda' \equiv g_A^2/(\Delta-\delta/2)=g_B^2/(\Delta+\delta/2)$, should satisfy
\begin{equation}\label{condition}
	4\lambda'^{2} \geq \delta^2.
\end{equation}
If the equal sign holds, the initial state $|gg4\rangle$ will evolve to the AMES, saying
\begin{equation}\label{eq_AMES_nr}
\begin{split}
	|\psi(T_e)\rangle=&~{\rm exp}(-iH_{\rm eff,mis}T_e)|gg4\rangle\\
	=&~\frac{1}{2}(-|ee1\rangle+|ge2\rangle-|eg3\rangle+|gg4\rangle),
\end{split}
\end{equation}
where the time interval $T_e=\pi/\sqrt{\delta^2+4\lambda'^2}$. Performing a single-qubit phase operation on qubit A, the minus signs of the first and third terms in Eq.~(\ref{eq_AMES_nr}) will be corrected. On the contrary, the entangled state prepared by the resonant interaction needs single-qubit operations twice to correct the additional phase, which may introduce more noise than the near-resonant case (see Appendix~\ref{AppB}). 
%================================================================================
\begin{figure}[h]
	\centering
	\includegraphics[width=\linewidth]{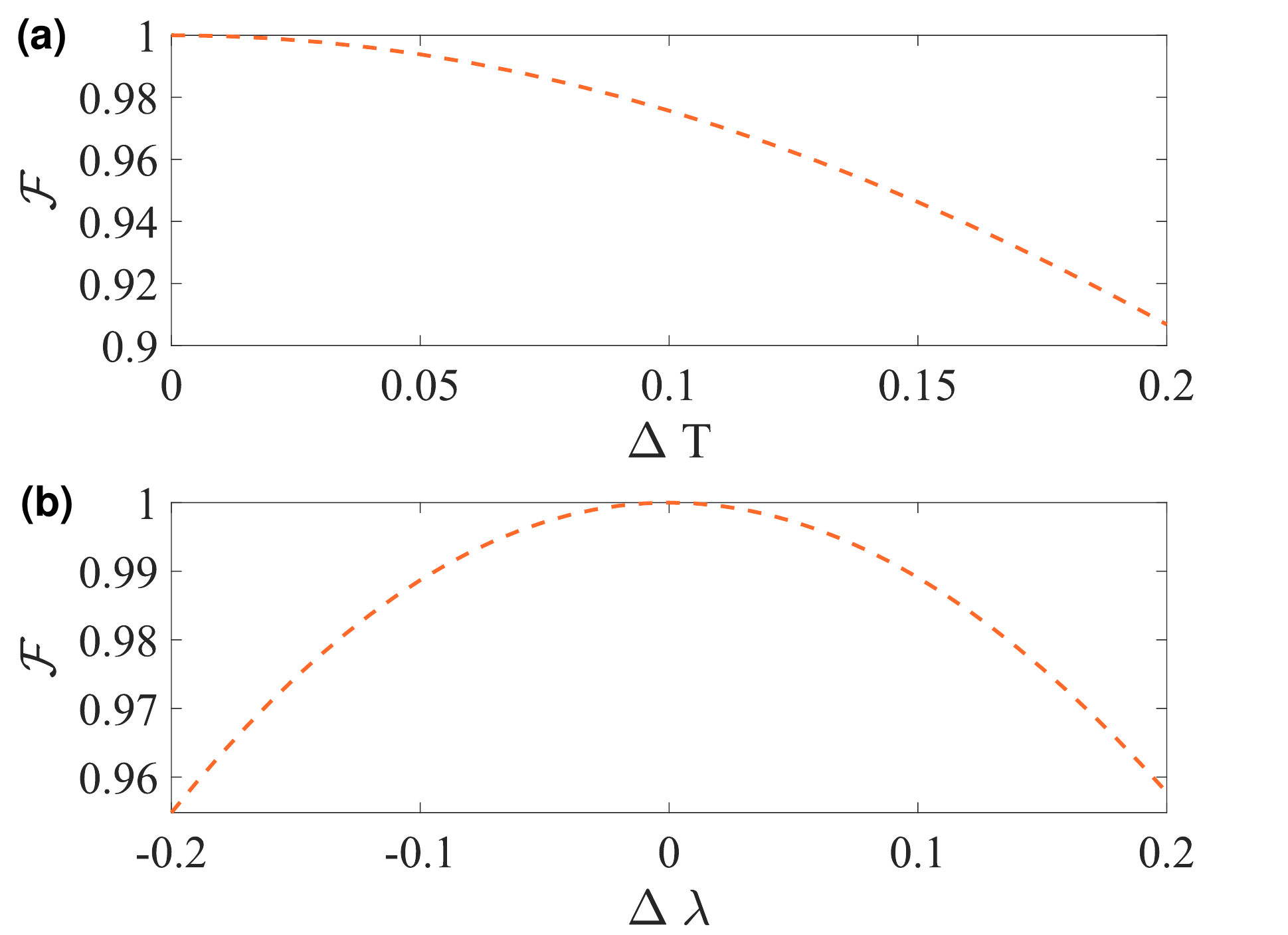}
	\captionsetup{justification=raggedright}
	\caption{The fidelity of the (2, 2, 4) state's preparation. We vary (a) the interaction time and (b) the coupling strength to compute the variation of the fidelity.}
	\label{fig.4}
\end{figure}
%================================================================================

In Fig.~\ref{fig.3}, we show the time evolution of the preparation of AMESs. Varying $|\lambda'/\delta|$, the evolution in Fig.~\ref{fig.3} (b), (c), and (d) are quite different. If the frequency mismatch exists, the maximal entanglement should be conditionally prepared. As we increase $\lambda'$ until it satisfies $4\lambda'^{2}= \delta^2$, the AMES will be prepared with the Rabi oscillation. In Fig.~\ref{fig.3} (d), the system is in the maximal entanglement at $t=\pi/\sqrt{\delta^2+4\lambda'^2}$, which is the extreme point of the time evolution. It also displays that the preparation of the maximally entangled state requires less temporal-control accuracy compared to the totally resonant scheme shown in Fig.~\ref{fig.3} (a).

Figure~\ref{fig.4} displays the fidelity versus error of time or the coupling strength. A temporal error will affect the fidelity of the preparation. The fidelity is still above $0.993$ as the error is less than $\pm0.05\times T_e$.
In addition, if the coupling strength has an error, i.e., $\pm 0.1 \times \lambda'$ but the interaction time is not changed, the fidelity is still above $0.989$. Therefore, one can generate the asymmetric maximally entangled state with high fidelity in this scheme, where the error of interaction time and coupling strength affects the fidelity slightly.

In experiments, the parameters discussed in Sec.~\ref {SecII} can also be used, e.g., the AMES is prepared after $\pi/4\lambda=8\times 10^ {-10} ~{\rm s} $ in the resonant case. With $\delta=2\pi\times 304~ {\rm MHz}$ and $\lambda'\approx 2\pi\times152~ {\rm MHz}$, satisfying Eq.~\ref{condition}, the AMES will be generated after $\pi/\sqrt{\delta^2+4\lambda'^2} =1.2\times 10^ {-9} ~{\rm s} $.
%================================================================================
%================================================================================

\section{Quantum Teleportation Between Two Qubits and a Ququart}\label{SecIV}

With the previous efficient preparation of the AMES, we can make use of this entanglement channel to implement quantum state teleportation for atomic qubits and an atomic ququart. Here, we introduce a quantum teleportation protocol based on the AMES~\cite{224_photon} to transfer quantum information initially encoded on the two-qubit system to a ququart. In this protocol, three particles of this entangled system stay away from others. We assume that Alice, Bob, and Charles hold qubit A, qubit B, and ququart C, respectively. Alice and Bob also hold qubits ${\rm d_1} $ and ${\rm d_2} $ of other qubits pair, whose unknown quantum state can be written as
\begin{equation}
\begin{split}
	|\phi\rangle_{\rm d_1,d_2}=&~c_1|gg\rangle_{\rm d_1,d_2}+c_2|ge\rangle_{\rm d_1,d_2}\\
	&~+c_3|eg\rangle_{\rm d_1,d_2}+c_4|ee\rangle_{\rm d_1,d_2}.
\end{split}
\end{equation}

Before measurements and operations, the combined state of all above five particles is
\begin{equation}
\begin{split}
	&~|\phi\rangle_{\rm d_1,d_2} \otimes |\psi_{2,2,4}\rangle_{\rm A,B,C}\\
	=&~|\Phi^+\rangle_{\rm d_1,A}|\Phi^+\rangle_{\rm d_2,B} (c_1|1\rangle_{\rm C}+c_2|2\rangle_{\rm C}+c_3|3\rangle_{\rm C}+c_4|4\rangle_{\rm C})\\
    &~+|\Phi^+\rangle_{\rm d_1,A}|\Phi^-\rangle_{\rm d_2,B} (c_1|1\rangle_{\rm C}-c_2|2\rangle_{\rm C}+c_3|3\rangle_{\rm C}-c_4|4\rangle_{\rm C})\\
	&~...\\
	&~+|\Psi^-\rangle_{\rm d_1,A}|\Psi^+\rangle_{\rm d_2,B} (-c_4|1\rangle_{\rm C}-c_3|2\rangle_{\rm C}+c_2|3\rangle_{\rm C}+c_1|4\rangle_{\rm C})\\
    &~+|\Psi^-\rangle_{\rm d_1,A}|\Psi^-\rangle_{\rm d_2,B} (c_4|1\rangle_{\rm C}-c_3|2\rangle_{\rm C}-c_2|3\rangle_{\rm C}+c_1|4\rangle_{\rm C}),
\end{split}
\end{equation}
where, $|\Phi^{\pm}\rangle=(|gg\rangle\pm|ee\rangle)/\sqrt{2}$, and $|\Psi^{\pm}\rangle=(|ge\rangle\pm|eg\rangle)/\sqrt{2}$ are the two-body Bell states. Thus, to teleport the unknown state, Alice will perform the Bell measurement on qubit A and ${\rm d_1}$, while Bob will also perform the Bell measurement on qubit B and ${\rm d_2}$. Their measurement result will send to Charles, and he will apply an appropriate 4-dimensional unitary operation to ququart C, and finally the quantum state of the ququart becomes
\begin{equation}
\begin{split}
    	|\phi\rangle_{\rm C}=&~c_1|1\rangle_{\rm C}+c_2|2\rangle_{\rm C}+c_3|3\rangle_{\rm C}+c_4|4\rangle_{\rm C}.
\end{split}
\end{equation}

We can also expand this scenario to teleport a single ququart information to a pair of distanced qubits via the (2, 2, 4) entangled state. If Charles holds a ququart with an unknown state's information and the ququart of the (2, 2, 4) state, he will perform the 4-dimensional Bell measurement and send the result to Alice with the classical communication. And Alice performs corresponding 2-dimensional unitary operations on her qubit A and qubit B of the (2, 2, 4) entangled state, thus the quantum state's information from Charles' ququart will be transferred to Alice's qubits.

In addition, the AEMS can also be used for realizing many other applications, such as remote state preparation and information compression. It could be a promising entanglement source in quantum networks and quantum computation for quantum objects in different dimensions.

%================================================================================
\begin{figure}
	\centering
	\includegraphics[width=\linewidth]{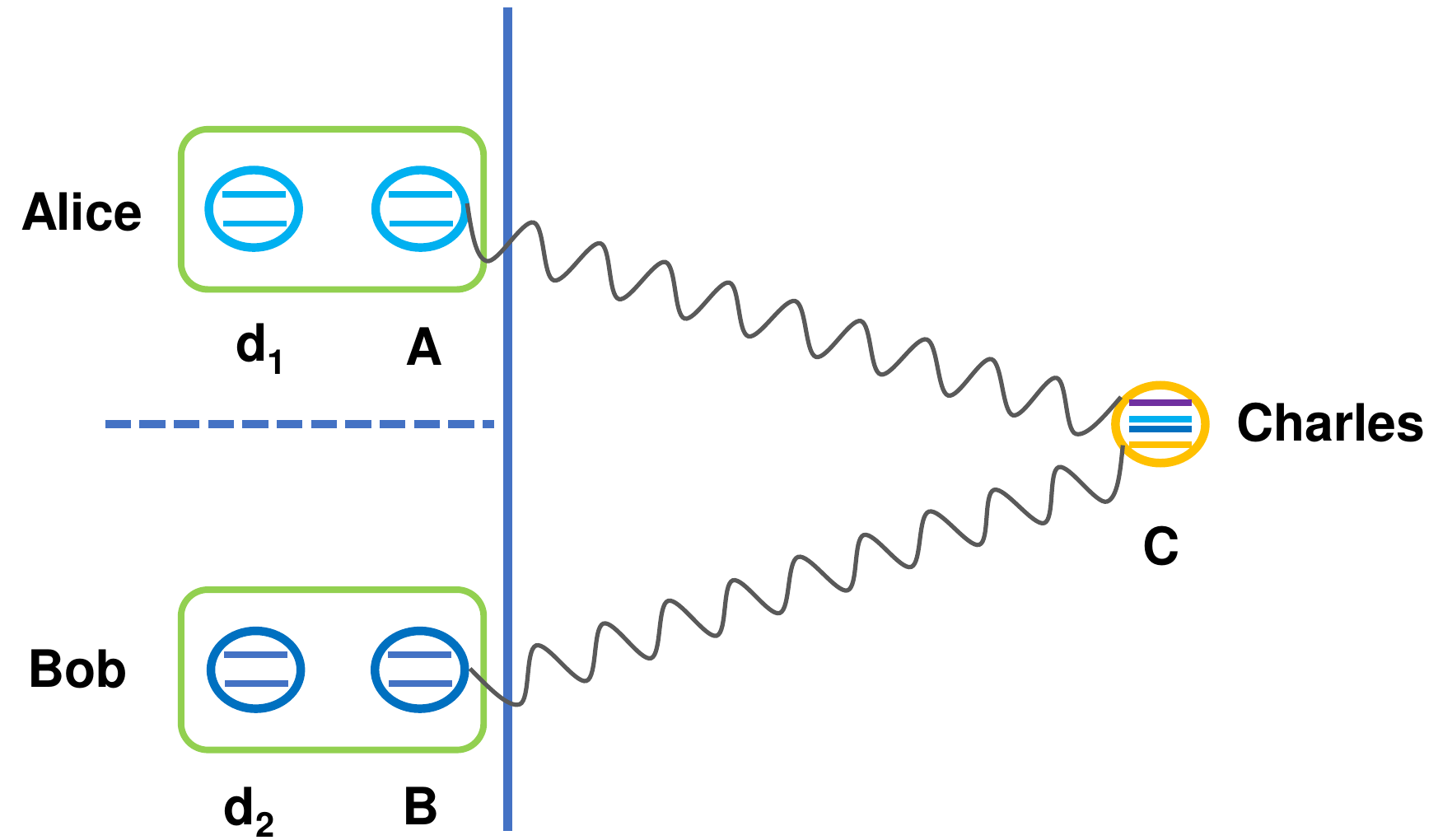}
	\captionsetup{justification=raggedright}
	\caption{Sketch of the quantum state teleportation from two far-separated qubits to a single ququart, and we utilize the asymmetric maximal entangled state prepared in the previous section as the communication channel.}
	\label{fig.5}
\end{figure}
%================================================================================

%================================================================================
%================================================================================

\section{Conclusion}\label{SecV}

In summary, we discuss the state transfer and the entanglement generation between 2- and 4-level atoms within a cavity to implement an information interface for quantum particles in different dimensions. We derive an effective qubit-ququart interaction with the far-detuning condition, which allows suppressed decoherence of the cavity. Our cavity QED scheme can obtain a perfect transfer process for arbitrary two-qubit quantum states and its inverse efficiently, without measurements and post selections. Moreover, the proposed scheme can be made use of determinately preparing the atomic asymmetric maximally entangled state, and we find there is a bound of atomic coupling strength for generating the AMES in high fidelity. This high-fidelity preparation can be applied in the remote state transfer between qubits and a ququart. The cavity QED would be a promising method to establish an interface for local and non-local information communication between particles with different dimensions.

As a prospect, a solid-state quantum storage protocol is expected to be implemented, where the information from multiple qubits can be stored in a multi-level atom. Specifically, through Raman transitions, i.e., a combination of classical driving fields and multi-mode photonic channels~\cite {PRL_state_transfer, raman_spin-controlled, multi-mode_cat-state, two-photon_laser}, multiple qubits' state will be transferred to multi-level atom's ground states, which allows a long lifetime of the storage. Multiple such high-dimensional atomic registers could be a promising platform for distributing quantum information in quantum networks. Furthermore, a non-local entanglement between qubits and qudits can be achieved in waveguide QED, where unidirectional coupling between quantum systems can prepare steady entangled states~\cite{driven-dissipative_preparation_entanglement,spin_networks}. This cascaded architecture is potential for realizing long-distance quantum state transfer in hybrid-dimensional systems.
%================================================================================
%================================================================================

\section{Acknowledgement}

This work is supported by the National Natural Science Foundation of China (Grant No. 11874432, 61974168), the National Key R\&D Program of China (Grant No. 2019YFA0308200, 2017YFA0305200), Special Project for Research and Development in Key Areas of Guangdong Province (2018B030329001, 2018B030325001).

\appendix

\section{Effective Hamiltonian of the Qubit-Ququart Cavity QED Model}\label{AppA}

In this part, we will give a detailed derivation of the effective atomic interaction in the main text. The original Hamiltonian is
\begin{equation}
\begin{split}
	H=&H_{1}+H_{2}+H_{3},\\
	H_{1}=&~\omega_{\rm op}a^{\dagger}a+\omega_{\rm at} \sigma_{ A}^{\dagger}\sigma_{ A}+ g_{ A}(\sigma_{ A}^{\dagger}a+a^{\dagger}\sigma_{ A}),\\
	H_{2}=&~\omega_{\rm op} b^{\dagger}b+\omega_{\rm at} \sigma_{ B}^{\dagger}\sigma_{ B}+ g_{ B}(\sigma_{ B}^{\dagger}b+b^{\dagger}\sigma_{ B}),\\
	H_{3}=&~\omega_{\rm op}(|2\rangle\langle 2|+|3\rangle \langle3| +2|4\rangle\langle4|)\\
	&~+g_{ A}[(|2\rangle\langle1|+|4\rangle\langle3|)a+a^{\dagger}(|1\rangle \langle2|+|3\rangle\langle4|)]\\
	&~+g_{ B}[(|3\rangle\langle1|+|4\rangle\langle2|)b+b^{\dagger}(|1\rangle \langle3|+|2\rangle\langle4|)],
\end{split}
\end{equation}
and the collective operators read
\begin{equation}
\begin{split}
	D_{ A}=\sigma_{ A}+|1\rangle \langle2|+|3\rangle\langle4|,\\
	D_{ B}=\sigma_{ B}+|1\rangle \langle3|+|2\rangle\langle4|.\\
\end{split}
\end{equation}
The Hamiltonian changes to
\begin{equation}\label{aeq_totalH}
\begin{split}
	H_0=&~\omega_{\rm op}(a^{\dagger}a+b^{\dagger}b)+\omega_{\rm op}(\sigma_{ A}^{\dagger}\sigma_{ A}+\sigma_{ B}^{\dagger}\sigma_{ B}+|2\rangle\langle 2|\\
    &~+|3\rangle \langle3| +2|4\rangle\langle4|),\\
    H_{\rm I}=&~g_{ A}(a^{\dagger}D_{ A}+D_{ A}^{\dagger}a)+g_{ B}(b^{\dagger}D_{ B}+D_{ B}^{\dagger}b).
\end{split}
\end{equation}
To obtain an effective second-order interaction between atoms, we consider the far-detuning regime, saying $\Delta=|\omega_{\rm op}-\omega_{\rm at}| \gg |g_{A/B}|$. In this case, the dynamics of cavity photons can be separated from the atomic system. Consider a unitary transformation $U={\rm exp}(S)$. The unitary transformation is expanded to
\begin{equation}
\begin{split}
	e^S (H_0+H_I) e^{-S}=&~H_0+H_I+[S,H_0]+[S,H_I]\\&~+
	\frac{1}{2}[S,[S,H_0]+\frac{1}{2}[S,[S,H_I]]+...
\end{split}
\end{equation}
To eliminate the first-order interaction with optical fields, the transformation satisfies
\begin{equation}\label{eq_SW}
\begin{split}
	H_I+[S,H_0]&=0\\
	U(H_0+H_I)U^{\dagger}&=H_0+\frac{1}{2}[S,H_I]+...
\end{split}
\end{equation}
In the far-detuning regime, the remaining terms can be neglected since they are high-order small values. $S$ is an anti-Hermitian operator and can be written as
\begin{equation}\label{eq_S}
	S=-\frac{g_A}{\Delta}(D_A^{\dagger}a-a^{\dagger}D_A)-\frac{g_B}{\Delta}(D_B^{\dagger}b-b^{\dagger}D_B).
\end{equation}
With Eq.~\eqref{eq_S}, we can compute the commutator of $S$ and $H_I$
\begin{equation}
\begin{split}
	&-\frac{g_A^2}{\Delta}[D_A^{\dagger}a-a^{\dagger}D_A,a^{\dagger}D_A+D_A^{\dagger}a]\\
	=&-\frac{2g_A^2}{\Delta}[D_A^{\dagger}D_A + (D_A^{\dagger}D_A-D_AD_A^{\dagger})a^{\dagger}a],
\end{split}
\end{equation}
and
\begin{equation}
\begin{split}
	&-\frac{g_B^2}{\Delta}[D_B^{\dagger}b-b^{\dagger}D_B,b^{\dagger}D_B+D_B^{\dagger}b]\\
	=&-\frac{2g_B^2}{\Delta}[D_B^{\dagger}D_B + (D_B^{\dagger}D_B-D_BD_B^{\dagger})b^{\dagger}b],
\end{split}
\end{equation}
where the commutative relationship, $[a,a^{\dagger}]=[b,b^{\dagger}]=1$, has been used. 
And other commutators obviously equal zero. Finally, we obtain the effective Hamiltonian without the particle exchange between atoms and photons,
\begin{equation}\label{aeq_eff_CQED}
\begin{split}
	H_{\rm eff}\approx&H_0+\frac{1}{2}[S,H_I]\\
	=&H_0-\frac{g_{ A}^2}{\Delta_{ A}}[D_{ A}^{\dagger}D_{ A} +D_{ A}^{ z} a^{\dagger}a]-\frac{g_{ B}^2}{\Delta_{ B}}[D_{ B}^{\dagger}D_{ B} +D_{ B}^{ z} b^{\dagger}b],
\end{split}
\end{equation}
where for simplicity, collective phase operators are given by
\begin{equation}
\begin{split}
	D_{ A}^{ z}\equiv&~\sigma_{ A}^{\dagger}\sigma_{ A}-\sigma_{ A}\sigma_{ A}^{\dagger}+\sigma_{44}-\sigma_{33}+\sigma_{22}-\sigma_{11}\\
 	=&D_A^{\dagger}D_A-D_AD_A^{\dagger}\\
	D_{ B}^{ z}\equiv&~\sigma_{ B}^{\dagger}\sigma_{ B}-\sigma_{ B}\sigma_{ B}^{\dagger}+\sigma_{44}-\sigma_{22}+\sigma_{33}-\sigma_{11}\\
	=&D_B^{\dagger}D_B-D_BD_B^{\dagger}.
\end{split}
\end{equation}
We consider that cavity field modes are initially in vacuum, saying
\begin{equation}\label{aeq_eff_CQED_np}
	H_{\rm eff,vac}=H_0 +\lambda(D_{ A}^{\dagger}D_{ A}+D_{ B}^{\dagger}D_{ B}),
\end{equation}
where $\lambda\equiv-g_{ A}^2/{\Delta_{ A}}=-g_{ B}^2/{\Delta_{ B}}$.

The initial atomic state is 
\begin{equation}\label{aeq_ini_state}\begin{split}
		|\psi_{0,1}\rangle = (C_{gg}|gg\rangle+C_{ge}|ge\rangle+C_{eg}|eg\rangle+C_{ee}|ee\rangle) \otimes |1\rangle \\
	\end{split}
\end{equation}
To transfer the information of the two-qubit state to the ququart, the Hamiltonian in Eq.~\eqref{aeq_eff_CQED_np} can be written as a matrix. In the double-excitation subspace which is consisting of states $|ee1\rangle$, $|ge2\rangle$, $|eg3\rangle$ and $|gg4\rangle$, the matrix can be written as 
\begin{equation}\label{mat_H_eff,vac}
	\begin{pmatrix}
		2\omega_{\rm at}+ 2\lambda & \lambda &~\lambda &0\\
		\lambda & 2\omega_{\rm at}+2\lambda &0 & \lambda\\
		\lambda & 0 & 2\omega_{\rm at}+2\lambda & \lambda\\
		0 & \lambda & \lambda & 2\omega_{\rm at}+2\lambda\\\end{pmatrix}.
\end{equation}
Diagonalize the Hamiltonian in Eq.~\eqref{mat_H_eff,vac}, and four orthogonal eigenstates are
\begin{equation}
\begin{split}
	|E_1\rangle&=\frac{1}{2}(|ee1\rangle -|eg2\rangle -|ge3\rangle + |gg4\rangle) \\
	|E_2\rangle&=\frac{1}{2}(-|ee1\rangle -|eg2\rangle +|ge3\rangle + |gg4\rangle) \\
	|E_3\rangle&=\frac{1}{2}(-|ee1\rangle +|eg2\rangle -|ge3\rangle + |gg4\rangle) \\
	|E_4\rangle&=\frac{1}{2}(|ee1\rangle +|eg2\rangle +|ge3\rangle + |gg4\rangle)
\end{split}
\end{equation}
with the eigenvalues $2\omega_{\rm at}$, $\omega_{\rm at}+2\lambda$, $\omega_{\rm at}+2\lambda$ and $\omega_{\rm at}+4\lambda$. The initial state $|gg4\rangle$ can be expanded in the basis of eigenstates:
\begin{equation}
	|ee1\rangle=\frac{1}{2}(|E_1\rangle+|E_2\rangle+|E_3\rangle+|E_4\rangle)
\end{equation}
Then the time-dependent evolution reads
\begin{equation}
\begin{split}
	&{\rm exp}(-iH_{\rm eff,vac}t)|ee1\rangle\\
	=&~\frac{1}{2}[|E_1\rangle+e^{-i 2\lambda t}(|E_2\rangle+|E_3\rangle)+e^{-i 4\lambda t}|E_4\rangle],
\end{split}
\end{equation}
where we have discarded the global phase. The time-dependent state vector with the original basis can be written as
\begin{equation}\label{vec_H_eff,vac}
	\begin{pmatrix}
		1-2e^{-i2\lambda t}+e^{-i4\lambda t}\\
		-1+e^{-i4\lambda t}\\
		-1+e^{-i4\lambda t}\\
		1+2e^{-i2\lambda t}+e^{-i4\lambda t}\\\end{pmatrix}.
\end{equation}
When $t=\pi/2\lambda$, $|ee1\rangle$ will turn to $|gg4\rangle$. For states $|eg1\rangle$ and $|ge1\rangle$, $H_{\rm eff,vac}$ becomes $2\times 2$ matrices, i.e.,
\begin{equation}\label{mat_H_eff,vac,2d}
	H_{\rm eff,vac,2d}=
	\begin{pmatrix}
		\omega_{\rm at}+\lambda & \lambda\\
		\lambda & \omega_{\rm at}+\lambda\\
	\end{pmatrix}.
\end{equation}
The evolution of state $|eg1\rangle$ is
\begin{equation}
\begin{split}
	&{\rm exp}(-iH_{\rm eff,vac,2d}t) |ge1\rangle\\
	=&~\frac{1}{2} [(e^{-i2\lambda t}+1)|ge1\rangle + (e^{-i2\lambda t}-1)|gg2\rangle].
\end{split}
\end{equation}
The state evolution with the local interaction is
\begin{equation}
\begin{split}
	|\psi_{1}(\frac{\pi}{2\lambda})\rangle &= {\rm exp}(-i H_{\rm eff,vac}\frac{\pi}{2\lambda})|\psi_{0,1}\rangle\\
	&= |gg\rangle \otimes (C_{gg}|1\rangle-C_{eg}|2\rangle-C_{ge}|3\rangle+C_{ee}|4\rangle),
\end{split}
\end{equation}
To correct the additional phase, one can apply twice single-qubit phase gates on qubit A and B, respectively, saying
\begin{equation}
\begin{split}
    |\psi_{0,1}'\rangle=&~(|g\rangle_A\langle g|_A-|e\rangle_A\langle e|_A)(|g\rangle_B\langle g|_B-|e\rangle_B\langle e|_B)|\psi_{0,1}\rangle\\
    =&~(C_{gg}|gg\rangle-C_{eg}|eg\rangle-C_{ge}|ge\rangle+C_{ee}|ee\rangle) \otimes |1\rangle,
\end{split}
\end{equation}
then the final state of the evolution reads
\begin{equation}
\begin{split}
    &~{\rm exp}(-i H_{\rm eff,vac}\frac{\pi}{2\lambda})|\psi_{0,1}'\rangle\\
    =&|gg\rangle \otimes (C_{gg}|1\rangle+ C_{eg}|2\rangle+C_{ge}|3\rangle+C_{ee}|4\rangle).
\end{split}
\end{equation}
On the other hand, the effective Hamiltonian with the photon-number-induced phase shift,
\begin{equation}\label{aeq_Htrans_2p}
	H_{\rm eff,ph}=H_0 +\lambda(D_{ A}^{\dagger}D_{ A}+D_{ B}^{\dagger}D_{ B}+D_A^z+D_B^z),
\end{equation}
and there initially exists a single photon in both optical field modes. 
The Hamiltonian in $H_{\rm eff,ph}$ can be written as a block matrix. The block matrices Hamiltonian expanded by the basis with $|ge1\rangle$ and $|gg2\rangle$ is
\begin{equation}\label{mat_H_eff,p}
	H_{\rm eff,ph,2d}=
	\begin{pmatrix}
		\omega_{\rm at}-\lambda & \lambda\\
		\lambda & \omega_{\rm at}-\lambda\\
	\end{pmatrix}.
\end{equation}
The state evolution is
\begin{equation}
\begin{split}
	&{\rm exp}(-iH_{\rm eff,ph,2d}t) |ge1\rangle\\
	=&~\frac{1}{2} [(e^{i2\lambda t}+1)|eg1\rangle + (-e^{i2\lambda t}+1)|gg2\rangle].
\end{split}
\end{equation}
The block matrix of $H_{\rm eff,ph}$ expanded by the double-excitation basis is obviously same as $H_{\rm eff,vac}$. In the other words, the photon-number-induced phase shift does not change the block matrix in the double-excitation subspace. 
Using these matrices, the perfect transfer can be achieved
\begin{equation}
\begin{split}
	|\psi'(\frac{\pi}{2\lambda})\rangle &= {\rm exp}(-i H_{\rm eff,ph}\frac{\pi}{2\lambda})|\psi_{0,1}\rangle\\
	&= |gg\rangle \otimes (C_{gg}|1\rangle+C_{eg}|2\rangle+C_{ge}|3\rangle+C_{ee}|4\rangle).
\end{split}
\end{equation}
%================================================================================
%================================================================================

\section{Near-Resonant Atomic Interaction for the (2, 2, 4) Entanglement Preparation}\label{AppB}
We assume that the four-level atom are initially excited to the state $|4\rangle$, after $t=\pi/4\lambda$, the atomic state becomes
\begin{equation}\label{aeq_AMES}
|\psi(\frac{\pi}{4\lambda})\rangle=\frac{1}{2}(-|ee1\rangle+i|ge2\rangle+i|eg3\rangle+|gg4\rangle)
\end{equation}
We apply single-qubit phase gates on the qubit atoms, and the correction phase equals $-\pi/2$, the gates read
\begin{equation}
\begin{split}
	U_{\rm phase}(-\frac{\pi}{2})=&|g\rangle \langle g| + e^{-i\frac{\pi}{2}} |e\rangle \langle e|\\
	=&|g\rangle \langle g| - i|e\rangle \langle e|
\end{split}
\end{equation}
After the phase correction, the state in Eq.~\eqref{aeq_AMES} becomes
\begin{equation}
	|\psi'(\frac{\pi}{4\lambda})\rangle=\frac{1}{2}(|ee1\rangle+|ge2\rangle+|eg3\rangle+|gg4\rangle)
\end{equation}
which is the asymmetric maximum entangled state.
To prepare the (2, 2, 4) AMES more steadily, as discussed in the main text, we introduce a mismatch $\delta$ between the exciton states of atoms, and the Hamiltonian changes to
\begin{equation}\label{beq_H}
\begin{split}
	H_{\rm mis}=&H_{\rm 0,mis}+H_{\rm I},\\
	H_{\rm 0,mis}=&~\omega_{ A}(\sigma_{ A}^{\dagger}\sigma_{ A}+|3\rangle \langle 3|+|4\rangle \langle 4|)+\omega_{\rm op} a^{\dagger}a \\
	+&~\omega_{ B}(\sigma_{ B}^{\dagger}\sigma_{ B}+|2\rangle \langle 2|+|4\rangle \langle 4|)+\omega_{\rm op}b^{\dagger}b,\\
	H_{\rm I}=&g_{ A}(a^{\dagger}D_{ A}+D_{ A}^{\dagger}a)+g_{ B}(b^{\dagger}D_{ B}+D_{ B}^{\dagger}b).\\
\end{split}
\end{equation}
   Here $\omega_A=\omega_{\rm at}-\delta/2$, $\omega_B=\omega_{\rm at}+\delta/2$, where $\delta \ll \Delta_{A/B}$. The generator of the Schrieffer-Wolf transformation becomes
\begin{equation}
\begin{split}
	S_{\rm mis}=&~\frac{g_A}{\Delta}(a^{\dagger}\sigma_A-\sigma_A^{\dagger}a)+\frac{g_B}{\Delta}(b^{\dagger}\sigma_B-\sigma_B^{\dagger}b)\\
    +&~\frac{g_A}{\Delta-\delta/2}[a^{\dagger}(|1\rangle \langle 2|+|3\rangle \langle 4|)-(|2\rangle \langle1|+|4\rangle \langle 3|)a]\\
    +&~\frac{g_B}{\Delta+\delta/2}[b^{\dagger}(|1\rangle \langle 3|+|2\rangle \langle 4|)-(|3\rangle \langle1|+|4\rangle \langle 2|)b].
\end{split}
\end{equation}
And $H_{\rm mis}$ becomes
\begin{equation}\label{beq_H_eff_nr}
\begin{split}
	H_{\rm eff,mis}=H_{\rm 0,mis}-\frac{g_A^2}{\Delta-\delta/2}D_A^{\dagger}D_A-\frac{g_B^2}{\Delta+\delta/2}D_B^{\dagger}D_B,
\end{split}
\end{equation}

We assume that $\lambda'\equiv g_A^2/(-\delta)=g_B^2/(\Delta+\delta)$. We rewrite $H_{\rm eff,mis}$ in Eq.~\eqref{beq_H_eff_nr} as a matrix with basis $|ee1\rangle$, $|ge2\rangle$, $|eg3\rangle$ and $|gg4\rangle$:
\begin{equation}\label{mat_H_eff,mis}
    	\begin{pmatrix}
		\omega_A+\omega_B+2\lambda' & \lambda' &\lambda' &0\\
		\lambda' & 2\omega_B+2\lambda' &0 & \lambda'\\
		\lambda' & 0 & 2\omega_A+2\lambda' & \lambda'\\
		0 & \lambda' & \lambda' & \omega_A+\omega_B+2\lambda'\\\end{pmatrix}.
\end{equation}
The eigenstates of Eq.~\eqref{beq_H_eff_nr} are
\begin{equation}
\begin{split}
	|E_1\rangle&=\frac{2\lambda'|ee1\rangle +(\Omega+\delta)|eg2\rangle +(\Omega-\delta)|ge3\rangle + 2 \lambda'|gg4\rangle}{\sqrt{2\Omega^2+2\delta^2+8\lambda'^2}} ,\\
	|E_2\rangle&=\frac{|ee1\rangle-|gg4\rangle}{\sqrt{2}},\\
	|E_3\rangle&=\frac{\delta |ee1\rangle-2\lambda'|eg2\rangle +2\lambda'|ge3\rangle +\delta |gg4\rangle}{\sqrt{8\lambda'^2+2\delta^2}},\\
	|E_4\rangle&=\frac{2\lambda'|ee1\rangle -(\Omega-\delta)|eg2\rangle -(\Omega+\delta)|ge3\rangle + 2 \lambda'|gg4\rangle}{\sqrt{2\Omega^2+2\delta^2+8\lambda'^2}}, 
\end{split}
\end{equation}
with the eigenvalues
\begin{equation}
\begin{split}
	E_1&=\omega_A+\omega_B+\sqrt{(\omega_A-\omega_B)^2+4\lambda'^2},\\
	E_2&=E_3=\omega_A+\omega_B,\\
	E_4&=\omega_A+\omega_B-\sqrt{(\omega_A-\omega_B)^2+4\lambda'^2},
\end{split}
\end{equation}
respectively, where
\begin{equation}
\begin{split}
	\Omega\equiv &~\sqrt{(\omega_A-\omega_B)^2+4\lambda'^2}\\
	=&~\sqrt{\delta^2+4\lambda'^2}.
\end{split}
\end{equation}

We can expand the initial state $|gg4\rangle$ with the eigenstate basis
\begin{equation}
|ee1\rangle=a|E_1\rangle+b|E_2\rangle+c|E_3\rangle+d|E_4\rangle,
\end{equation}
where
\begin{equation}
\begin{split}
	a&=\frac{\lambda'\sqrt{2\Omega^2+2\delta^2+8\lambda'^2}}{2\Omega^2},
	~b=-\frac{1}{\sqrt{2}},\\
	c&=\frac{\delta\sqrt{8\lambda'^2+2\delta^2}}{2\lambda'^2},~d=\frac{\lambda'\sqrt{2\Omega^2+2\delta^2+8\lambda'^2}}{2\Omega^2}.
\end{split}
\end{equation}
Finally, the time-dependent evolution reads
\begin{equation}
\begin{split}
	|\psi(t)\rangle=&~{\rm exp}(-iH_{\rm eff,mis}t)|ee1\rangle\\
	=&ae^{-i(\omega_A+\omega_B+\Omega)t}|E_1\rangle+e^{-i(\omega_A+\omega_B)t}(b|E_2\rangle+c|E_3\rangle)\\
    &~+de^{-i(\omega_A+\omega_B-\Omega)t}|E_4\rangle
\end{split}
\end{equation}
To obtain the AMES, there exists a moment $t_e$ which satisfies
\begin{equation}
	(\langle ee1|+ \langle gg4|){\rm exp}(-iH_{\rm eff,mis}t_e)|ee1\rangle =0,
\end{equation}
and that means
\begin{equation}
	\cos\Omega t_e=-\frac{\delta^2}{4\gamma^2}.
\end{equation}
We obtain the condition to prepare the AMES 
\begin{equation}
	4\lambda'^2\geq \delta^2.
\end{equation}
If the equal sign holds, the AMES can be prepared after $T_e=\pi/\Omega$
\begin{equation}\label{aeq_AMES_nr}
	\frac{1}{2}(-|ee1\rangle+|ge2\rangle-|eg3\rangle+|gg4\rangle).
\end{equation}
With the single-qubit phase operation on the first qubit, qubit A, saying
\begin{equation}
\begin{split}
	U_{\rm phase}(\pi)=&|g\rangle_A \langle g|_A + e^{i\pi} |e\rangle_A \langle e|_A\\
	=&|g\rangle_A \langle g| _A- |e\rangle_A \langle e|_A,
\end{split}
\end{equation}
the AMES becomes
\begin{equation}
	|\psi_{2,2,4}\rangle=\frac{1}{2}(|ee1\rangle+|ge2\rangle+|eg3\rangle+|gg4\rangle)
\end{equation}

\bibliography{reference}

\begin{thebibliography}{73}%
\makeatletter
\providecommand \@ifxundefined [1]{%
 \@ifx{#1\undefined}
}%
\providecommand \@ifnum [1]{%
 \ifnum #1\expandafter \@firstoftwo
 \else \expandafter \@secondoftwo
 \fi
}%
\providecommand \@ifx [1]{%
 \ifx #1\expandafter \@firstoftwo
 \else \expandafter \@secondoftwo
 \fi
}%
\providecommand \natexlab [1]{#1}%
\providecommand \enquote  [1]{``#1''}%
\providecommand \bibnamefont  [1]{#1}%
\providecommand \bibfnamefont [1]{#1}%
\providecommand \citenamefont [1]{#1}%
\providecommand \href@noop [0]{\@secondoftwo}%
\providecommand \href [0]{\begingroup \@sanitize@url \@href}%
\providecommand \@href[1]{\@@startlink{#1}\@@href}%
\providecommand \@@href[1]{\endgroup#1\@@endlink}%
\providecommand \@sanitize@url [0]{\catcode `\\12\catcode `\$12\catcode
  `\&12\catcode `\#12\catcode `\^12\catcode `\_12\catcode `\%12\relax}%
\providecommand \@@startlink[1]{}%
\providecommand \@@endlink[0]{}%
\providecommand \url  [0]{\begingroup\@sanitize@url \@url }%
\providecommand \@url [1]{\endgroup\@href {#1}{\urlprefix }}%
\providecommand \urlprefix  [0]{URL }%
\providecommand \Eprint [0]{\href }%
\providecommand \doibase [0]{https://doi.org/}%
\providecommand \selectlanguage [0]{\@gobble}%
\providecommand \bibinfo  [0]{\@secondoftwo}%
\providecommand \bibfield  [0]{\@secondoftwo}%
\providecommand \translation [1]{[#1]}%
\providecommand \BibitemOpen [0]{}%
\providecommand \bibitemStop [0]{}%
\providecommand \bibitemNoStop [0]{.\EOS\space}%
\providecommand \EOS [0]{\spacefactor3000\relax}%
\providecommand \BibitemShut  [1]{\csname bibitem#1\endcsname}%
\let\auto@bib@innerbib\@empty
%</preamble>
\bibitem [{\citenamefont {O’Leary}\ \emph {et~al.}(2006)\citenamefont
  {O’Leary}, \citenamefont {Brennen},\ and\ \citenamefont
  {Bullock}}]{parallelism_atom_qudit}%
  \BibitemOpen
  \bibfield  {author} {\bibinfo {author} {\bibfnamefont {D.~P.}\ \bibnamefont
  {O’Leary}}, \bibinfo {author} {\bibfnamefont {G.~K.}\ \bibnamefont
  {Brennen}},\ and\ \bibinfo {author} {\bibfnamefont {S.~S.}\ \bibnamefont
  {Bullock}},\ }\bibfield  {title} {\bibinfo {title} {Parallelism for quantum
  computation with qudits},\ }\href@noop {} {\bibfield  {journal} {\bibinfo
  {journal} {Physical Review A}\ }\textbf {\bibinfo {volume} {74}},\ \bibinfo
  {pages} {032334} (\bibinfo {year} {2006})}\BibitemShut {NoStop}%
\bibitem [{\citenamefont {Godfrin}\ \emph {et~al.}(2017)\citenamefont
  {Godfrin}, \citenamefont {Ferhat}, \citenamefont {Ballou}, \citenamefont
  {Klyatskaya}, \citenamefont {Ruben}, \citenamefont {Wernsdorfer},\ and\
  \citenamefont {Balestro}}]{quantum_search_atom_qudit}%
  \BibitemOpen
  \bibfield  {author} {\bibinfo {author} {\bibfnamefont {C.}~\bibnamefont
  {Godfrin}}, \bibinfo {author} {\bibfnamefont {A.}~\bibnamefont {Ferhat}},
  \bibinfo {author} {\bibfnamefont {R.}~\bibnamefont {Ballou}}, \bibinfo
  {author} {\bibfnamefont {S.}~\bibnamefont {Klyatskaya}}, \bibinfo {author}
  {\bibfnamefont {M.}~\bibnamefont {Ruben}}, \bibinfo {author} {\bibfnamefont
  {W.}~\bibnamefont {Wernsdorfer}},\ and\ \bibinfo {author} {\bibfnamefont
  {F.}~\bibnamefont {Balestro}},\ }\bibfield  {title} {\bibinfo {title}
  {Operating quantum states in single magnetic molecules: Implementation of
  grover's quantum algorithm},\ }\href
  {https://doi.org/10.1103/PhysRevLett.119.187702} {\bibfield  {journal}
  {\bibinfo  {journal} {Phys. Rev. Lett.}\ }\textbf {\bibinfo {volume} {119}},\
  \bibinfo {pages} {187702} (\bibinfo {year} {2017})}\BibitemShut {NoStop}%
\bibitem [{\citenamefont {Calixto}\ \emph {et~al.}(2021)\citenamefont
  {Calixto}, \citenamefont {Mayorgas},\ and\ \citenamefont
  {Guerrero}}]{entanglement_squeezing_D-level}%
  \BibitemOpen
  \bibfield  {author} {\bibinfo {author} {\bibfnamefont {M.}~\bibnamefont
  {Calixto}}, \bibinfo {author} {\bibfnamefont {A.}~\bibnamefont {Mayorgas}},\
  and\ \bibinfo {author} {\bibfnamefont {J.}~\bibnamefont {Guerrero}},\
  }\bibfield  {title} {\bibinfo {title} {Entanglement and u (d)-spin squeezing
  in symmetric multi-qudit systems and applications to quantum phase
  transitions in lipkin--meshkov--glick d-level atom models},\ }\href@noop {}
  {\bibfield  {journal} {\bibinfo  {journal} {Quantum Information Processing}\
  }\textbf {\bibinfo {volume} {20}},\ \bibinfo {pages} {1} (\bibinfo {year}
  {2021})}\BibitemShut {NoStop}%
\bibitem [{\citenamefont {Klimov}\ \emph {et~al.}(2003)\citenamefont {Klimov},
  \citenamefont {Guzm\'an}, \citenamefont {Retamal},\ and\ \citenamefont
  {Saavedra}}]{qutrit_simulation}%
  \BibitemOpen
  \bibfield  {author} {\bibinfo {author} {\bibfnamefont {A.~B.}\ \bibnamefont
  {Klimov}}, \bibinfo {author} {\bibfnamefont {R.}~\bibnamefont {Guzm\'an}},
  \bibinfo {author} {\bibfnamefont {J.~C.}\ \bibnamefont {Retamal}},\ and\
  \bibinfo {author} {\bibfnamefont {C.}~\bibnamefont {Saavedra}},\ }\bibfield
  {title} {\bibinfo {title} {Qutrit quantum computer with trapped ions},\
  }\href {https://doi.org/10.1103/PhysRevA.67.062313} {\bibfield  {journal}
  {\bibinfo  {journal} {Phys. Rev. A}\ }\textbf {\bibinfo {volume} {67}},\
  \bibinfo {pages} {062313} (\bibinfo {year} {2003})}\BibitemShut {NoStop}%
\bibitem [{\citenamefont {Ringbauer}\ \emph {et~al.}(2022)\citenamefont
  {Ringbauer}, \citenamefont {Meth}, \citenamefont {Postler}, \citenamefont
  {Stricker}, \citenamefont {Blatt}, \citenamefont {Schindler},\ and\
  \citenamefont {Monz}}]{qudit_ion}%
  \BibitemOpen
  \bibfield  {author} {\bibinfo {author} {\bibfnamefont {M.}~\bibnamefont
  {Ringbauer}}, \bibinfo {author} {\bibfnamefont {M.}~\bibnamefont {Meth}},
  \bibinfo {author} {\bibfnamefont {L.}~\bibnamefont {Postler}}, \bibinfo
  {author} {\bibfnamefont {R.}~\bibnamefont {Stricker}}, \bibinfo {author}
  {\bibfnamefont {R.}~\bibnamefont {Blatt}}, \bibinfo {author} {\bibfnamefont
  {P.}~\bibnamefont {Schindler}},\ and\ \bibinfo {author} {\bibfnamefont
  {T.}~\bibnamefont {Monz}},\ }\bibfield  {title} {\bibinfo {title} {A
  universal qudit quantum processor with trapped ions},\ }\href@noop {}
  {\bibfield  {journal} {\bibinfo  {journal} {Nature Physics}\ ,\ \bibinfo
  {pages} {1}} (\bibinfo {year} {2022})}\BibitemShut {NoStop}%
\bibitem [{\citenamefont {Sparrow}\ \emph {et~al.}(2018)\citenamefont
  {Sparrow}, \citenamefont {Mart{\'\i}n-L{\'o}pez}, \citenamefont {Maraviglia},
  \citenamefont {Neville}, \citenamefont {Harrold}, \citenamefont {Carolan},
  \citenamefont {Joglekar}, \citenamefont {Hashimoto}, \citenamefont {Matsuda},
  \citenamefont {O’Brien} \emph {et~al.}}]{photonic_molecules_vibration}%
  \BibitemOpen
  \bibfield  {author} {\bibinfo {author} {\bibfnamefont {C.}~\bibnamefont
  {Sparrow}}, \bibinfo {author} {\bibfnamefont {E.}~\bibnamefont
  {Mart{\'\i}n-L{\'o}pez}}, \bibinfo {author} {\bibfnamefont {N.}~\bibnamefont
  {Maraviglia}}, \bibinfo {author} {\bibfnamefont {A.}~\bibnamefont {Neville}},
  \bibinfo {author} {\bibfnamefont {C.}~\bibnamefont {Harrold}}, \bibinfo
  {author} {\bibfnamefont {J.}~\bibnamefont {Carolan}}, \bibinfo {author}
  {\bibfnamefont {Y.~N.}\ \bibnamefont {Joglekar}}, \bibinfo {author}
  {\bibfnamefont {T.}~\bibnamefont {Hashimoto}}, \bibinfo {author}
  {\bibfnamefont {N.}~\bibnamefont {Matsuda}}, \bibinfo {author} {\bibfnamefont
  {J.~L.}\ \bibnamefont {O’Brien}}, \emph {et~al.},\ }\bibfield  {title}
  {\bibinfo {title} {Simulating the vibrational quantum dynamics of molecules
  using photonics},\ }\href@noop {} {\bibfield  {journal} {\bibinfo  {journal}
  {Nature}\ }\textbf {\bibinfo {volume} {557}},\ \bibinfo {pages} {660}
  (\bibinfo {year} {2018})}\BibitemShut {NoStop}%
\bibitem [{\citenamefont {Chi}\ \emph {et~al.}(2022)\citenamefont {Chi},
  \citenamefont {Huang}, \citenamefont {Zhang}, \citenamefont {Mao},
  \citenamefont {Zhou}, \citenamefont {Chen}, \citenamefont {Zhai},
  \citenamefont {Bao}, \citenamefont {Dai}, \citenamefont {Yuan} \emph
  {et~al.}}]{qudit-based_processer_photon}%
  \BibitemOpen
  \bibfield  {author} {\bibinfo {author} {\bibfnamefont {Y.}~\bibnamefont
  {Chi}}, \bibinfo {author} {\bibfnamefont {J.}~\bibnamefont {Huang}}, \bibinfo
  {author} {\bibfnamefont {Z.}~\bibnamefont {Zhang}}, \bibinfo {author}
  {\bibfnamefont {J.}~\bibnamefont {Mao}}, \bibinfo {author} {\bibfnamefont
  {Z.}~\bibnamefont {Zhou}}, \bibinfo {author} {\bibfnamefont {X.}~\bibnamefont
  {Chen}}, \bibinfo {author} {\bibfnamefont {C.}~\bibnamefont {Zhai}}, \bibinfo
  {author} {\bibfnamefont {J.}~\bibnamefont {Bao}}, \bibinfo {author}
  {\bibfnamefont {T.}~\bibnamefont {Dai}}, \bibinfo {author} {\bibfnamefont
  {H.}~\bibnamefont {Yuan}}, \emph {et~al.},\ }\bibfield  {title} {\bibinfo
  {title} {A programmable qudit-based quantum processor},\ }\href@noop {}
  {\bibfield  {journal} {\bibinfo  {journal} {Nature communications}\ }\textbf
  {\bibinfo {volume} {13}},\ \bibinfo {pages} {1} (\bibinfo {year}
  {2022})}\BibitemShut {NoStop}%
\bibitem [{\citenamefont {Giordani}\ \emph {et~al.}(2019)\citenamefont
  {Giordani}, \citenamefont {Polino}, \citenamefont {Emiliani}, \citenamefont
  {Suprano}, \citenamefont {Innocenti}, \citenamefont {Majury}, \citenamefont
  {Marrucci}, \citenamefont {Paternostro}, \citenamefont {Ferraro},
  \citenamefont {Spagnolo} \emph {et~al.}}]{qudit_photons_quantum-walks}%
  \BibitemOpen
  \bibfield  {author} {\bibinfo {author} {\bibfnamefont {T.}~\bibnamefont
  {Giordani}}, \bibinfo {author} {\bibfnamefont {E.}~\bibnamefont {Polino}},
  \bibinfo {author} {\bibfnamefont {S.}~\bibnamefont {Emiliani}}, \bibinfo
  {author} {\bibfnamefont {A.}~\bibnamefont {Suprano}}, \bibinfo {author}
  {\bibfnamefont {L.}~\bibnamefont {Innocenti}}, \bibinfo {author}
  {\bibfnamefont {H.}~\bibnamefont {Majury}}, \bibinfo {author} {\bibfnamefont
  {L.}~\bibnamefont {Marrucci}}, \bibinfo {author} {\bibfnamefont
  {M.}~\bibnamefont {Paternostro}}, \bibinfo {author} {\bibfnamefont
  {A.}~\bibnamefont {Ferraro}}, \bibinfo {author} {\bibfnamefont
  {N.}~\bibnamefont {Spagnolo}}, \emph {et~al.},\ }\bibfield  {title} {\bibinfo
  {title} {Experimental engineering of arbitrary qudit states with
  discrete-time quantum walks},\ }\href@noop {} {\bibfield  {journal} {\bibinfo
   {journal} {Physical review letters}\ }\textbf {\bibinfo {volume} {122}},\
  \bibinfo {pages} {020503} (\bibinfo {year} {2019})}\BibitemShut {NoStop}%
\bibitem [{\citenamefont {Niu}\ \emph {et~al.}(2018)\citenamefont {Niu},
  \citenamefont {Chuang},\ and\ \citenamefont
  {Shapiro}}]{qudit-based_universal_computation}%
  \BibitemOpen
  \bibfield  {author} {\bibinfo {author} {\bibfnamefont {M.~Y.}\ \bibnamefont
  {Niu}}, \bibinfo {author} {\bibfnamefont {I.~L.}\ \bibnamefont {Chuang}},\
  and\ \bibinfo {author} {\bibfnamefont {J.~H.}\ \bibnamefont {Shapiro}},\
  }\bibfield  {title} {\bibinfo {title} {Qudit-basis universal quantum
  computation using ${\ensuremath{\chi}}^{(2)}$ interactions},\ }\href
  {https://doi.org/10.1103/PhysRevLett.120.160502} {\bibfield  {journal}
  {\bibinfo  {journal} {Phys. Rev. Lett.}\ }\textbf {\bibinfo {volume} {120}},\
  \bibinfo {pages} {160502} (\bibinfo {year} {2018})}\BibitemShut {NoStop}%
\bibitem [{\citenamefont {Neeley}\ \emph {et~al.}(2009)\citenamefont {Neeley},
  \citenamefont {Ansmann}, \citenamefont {Bialczak}, \citenamefont {Hofheinz},
  \citenamefont {Lucero}, \citenamefont {O'Connell}, \citenamefont {Sank},
  \citenamefont {Wang}, \citenamefont {Wenner}, \citenamefont {Cleland},
  \citenamefont {Geller},\ and\ \citenamefont
  {Martinis}}]{Emulation_Superconducting_qudit}%
  \BibitemOpen
  \bibfield  {author} {\bibinfo {author} {\bibfnamefont {M.}~\bibnamefont
  {Neeley}}, \bibinfo {author} {\bibfnamefont {M.}~\bibnamefont {Ansmann}},
  \bibinfo {author} {\bibfnamefont {R.~C.}\ \bibnamefont {Bialczak}}, \bibinfo
  {author} {\bibfnamefont {M.}~\bibnamefont {Hofheinz}}, \bibinfo {author}
  {\bibfnamefont {E.}~\bibnamefont {Lucero}}, \bibinfo {author} {\bibfnamefont
  {A.~D.}\ \bibnamefont {O'Connell}}, \bibinfo {author} {\bibfnamefont
  {D.}~\bibnamefont {Sank}}, \bibinfo {author} {\bibfnamefont {H.}~\bibnamefont
  {Wang}}, \bibinfo {author} {\bibfnamefont {J.}~\bibnamefont {Wenner}},
  \bibinfo {author} {\bibfnamefont {A.~N.}\ \bibnamefont {Cleland}}, \bibinfo
  {author} {\bibfnamefont {M.~R.}\ \bibnamefont {Geller}},\ and\ \bibinfo
  {author} {\bibfnamefont {J.~M.}\ \bibnamefont {Martinis}},\ }\bibfield
  {title} {\bibinfo {title} {Emulation of a quantum spin with a superconducting
  phase qudit},\ }\href {https://doi.org/10.1126/science.1173440} {\bibfield
  {journal} {\bibinfo  {journal} {Science}\ }\textbf {\bibinfo {volume}
  {325}},\ \bibinfo {pages} {722} (\bibinfo {year} {2009})},\ \Eprint
  {https://arxiv.org/abs/https://www.science.org/doi/pdf/10.1126/science.1173440}
  {https://www.science.org/doi/pdf/10.1126/science.1173440} \BibitemShut
  {NoStop}%
\bibitem [{\citenamefont {Blok}\ \emph {et~al.}(2021)\citenamefont {Blok},
  \citenamefont {Ramasesh}, \citenamefont {Schuster}, \citenamefont {O'Brien},
  \citenamefont {Kreikebaum}, \citenamefont {Dahlen}, \citenamefont {Morvan},
  \citenamefont {Yoshida}, \citenamefont {Yao},\ and\ \citenamefont
  {Siddiqi}}]{Scrambling_superconducting_qutrit}%
  \BibitemOpen
  \bibfield  {author} {\bibinfo {author} {\bibfnamefont {M.~S.}\ \bibnamefont
  {Blok}}, \bibinfo {author} {\bibfnamefont {V.~V.}\ \bibnamefont {Ramasesh}},
  \bibinfo {author} {\bibfnamefont {T.}~\bibnamefont {Schuster}}, \bibinfo
  {author} {\bibfnamefont {K.}~\bibnamefont {O'Brien}}, \bibinfo {author}
  {\bibfnamefont {J.~M.}\ \bibnamefont {Kreikebaum}}, \bibinfo {author}
  {\bibfnamefont {D.}~\bibnamefont {Dahlen}}, \bibinfo {author} {\bibfnamefont
  {A.}~\bibnamefont {Morvan}}, \bibinfo {author} {\bibfnamefont
  {B.}~\bibnamefont {Yoshida}}, \bibinfo {author} {\bibfnamefont {N.~Y.}\
  \bibnamefont {Yao}},\ and\ \bibinfo {author} {\bibfnamefont {I.}~\bibnamefont
  {Siddiqi}},\ }\bibfield  {title} {\bibinfo {title} {Quantum information
  scrambling on a superconducting qutrit processor},\ }\href
  {https://doi.org/10.1103/PhysRevX.11.021010} {\bibfield  {journal} {\bibinfo
  {journal} {Phys. Rev. X}\ }\textbf {\bibinfo {volume} {11}},\ \bibinfo
  {pages} {021010} (\bibinfo {year} {2021})}\BibitemShut {NoStop}%
\bibitem [{\citenamefont {Soltamov}\ \emph {et~al.}(2019)\citenamefont
  {Soltamov}, \citenamefont {Kasper}, \citenamefont {Poshakinskiy},
  \citenamefont {Anisimov}, \citenamefont {Mokhov}, \citenamefont {Sperlich},
  \citenamefont {Tarasenko}, \citenamefont {Baranov}, \citenamefont
  {Astakhov},\ and\ \citenamefont {Dyakonov}}]{spin-qudit_silicon_carbide}%
  \BibitemOpen
  \bibfield  {author} {\bibinfo {author} {\bibfnamefont {V.}~\bibnamefont
  {Soltamov}}, \bibinfo {author} {\bibfnamefont {C.}~\bibnamefont {Kasper}},
  \bibinfo {author} {\bibfnamefont {A.}~\bibnamefont {Poshakinskiy}}, \bibinfo
  {author} {\bibfnamefont {A.}~\bibnamefont {Anisimov}}, \bibinfo {author}
  {\bibfnamefont {E.}~\bibnamefont {Mokhov}}, \bibinfo {author} {\bibfnamefont
  {A.}~\bibnamefont {Sperlich}}, \bibinfo {author} {\bibfnamefont
  {S.}~\bibnamefont {Tarasenko}}, \bibinfo {author} {\bibfnamefont
  {P.}~\bibnamefont {Baranov}}, \bibinfo {author} {\bibfnamefont
  {G.}~\bibnamefont {Astakhov}},\ and\ \bibinfo {author} {\bibfnamefont
  {V.}~\bibnamefont {Dyakonov}},\ }\bibfield  {title} {\bibinfo {title}
  {Excitation and coherent control of spin qudit modes in silicon carbide at
  room temperature},\ }\href@noop {} {\bibfield  {journal} {\bibinfo  {journal}
  {Nature communications}\ }\textbf {\bibinfo {volume} {10}},\ \bibinfo {pages}
  {1} (\bibinfo {year} {2019})}\BibitemShut {NoStop}%
\bibitem [{\citenamefont {Moreno-Pineda}\ \emph {et~al.}(2018)\citenamefont
  {Moreno-Pineda}, \citenamefont {Godfrin}, \citenamefont {Balestro},
  \citenamefont {Wernsdorfer},\ and\ \citenamefont
  {Ruben}}]{spin-qudit_molecular}%
  \BibitemOpen
  \bibfield  {author} {\bibinfo {author} {\bibfnamefont {E.}~\bibnamefont
  {Moreno-Pineda}}, \bibinfo {author} {\bibfnamefont {C.}~\bibnamefont
  {Godfrin}}, \bibinfo {author} {\bibfnamefont {F.}~\bibnamefont {Balestro}},
  \bibinfo {author} {\bibfnamefont {W.}~\bibnamefont {Wernsdorfer}},\ and\
  \bibinfo {author} {\bibfnamefont {M.}~\bibnamefont {Ruben}},\ }\bibfield
  {title} {\bibinfo {title} {Molecular spin qudits for quantum algorithms},\
  }\href@noop {} {\bibfield  {journal} {\bibinfo  {journal} {Chemical Society
  Reviews}\ }\textbf {\bibinfo {volume} {47}},\ \bibinfo {pages} {501}
  (\bibinfo {year} {2018})}\BibitemShut {NoStop}%
\bibitem [{\citenamefont {Bullock}\ \emph {et~al.}(2005)\citenamefont
  {Bullock}, \citenamefont {O'Leary},\ and\ \citenamefont
  {Brennen}}]{optimal_d-level}%
  \BibitemOpen
  \bibfield  {author} {\bibinfo {author} {\bibfnamefont {S.~S.}\ \bibnamefont
  {Bullock}}, \bibinfo {author} {\bibfnamefont {D.~P.}\ \bibnamefont
  {O'Leary}},\ and\ \bibinfo {author} {\bibfnamefont {G.~K.}\ \bibnamefont
  {Brennen}},\ }\bibfield  {title} {\bibinfo {title} {Asymptotically optimal
  quantum circuits for $d$-level systems},\ }\href
  {https://doi.org/10.1103/PhysRevLett.94.230502} {\bibfield  {journal}
  {\bibinfo  {journal} {Phys. Rev. Lett.}\ }\textbf {\bibinfo {volume} {94}},\
  \bibinfo {pages} {230502} (\bibinfo {year} {2005})}\BibitemShut {NoStop}%
\bibitem [{\citenamefont {Lanyon}\ \emph {et~al.}(2009)\citenamefont {Lanyon},
  \citenamefont {Barbieri}, \citenamefont {Almeida}, \citenamefont {Jennewein},
  \citenamefont {Ralph}, \citenamefont {Resch}, \citenamefont {Pryde},
  \citenamefont {O’brien}, \citenamefont {Gilchrist},\ and\ \citenamefont
  {White}}]{simplify_higher-dimensional}%
  \BibitemOpen
  \bibfield  {author} {\bibinfo {author} {\bibfnamefont {B.~P.}\ \bibnamefont
  {Lanyon}}, \bibinfo {author} {\bibfnamefont {M.}~\bibnamefont {Barbieri}},
  \bibinfo {author} {\bibfnamefont {M.~P.}\ \bibnamefont {Almeida}}, \bibinfo
  {author} {\bibfnamefont {T.}~\bibnamefont {Jennewein}}, \bibinfo {author}
  {\bibfnamefont {T.~C.}\ \bibnamefont {Ralph}}, \bibinfo {author}
  {\bibfnamefont {K.~J.}\ \bibnamefont {Resch}}, \bibinfo {author}
  {\bibfnamefont {G.~J.}\ \bibnamefont {Pryde}}, \bibinfo {author}
  {\bibfnamefont {J.~L.}\ \bibnamefont {O’brien}}, \bibinfo {author}
  {\bibfnamefont {A.}~\bibnamefont {Gilchrist}},\ and\ \bibinfo {author}
  {\bibfnamefont {A.~G.}\ \bibnamefont {White}},\ }\bibfield  {title} {\bibinfo
  {title} {Simplifying quantum logic using higher-dimensional hilbert spaces},\
  }\href@noop {} {\bibfield  {journal} {\bibinfo  {journal} {Nature Physics}\
  }\textbf {\bibinfo {volume} {5}},\ \bibinfo {pages} {134} (\bibinfo {year}
  {2009})}\BibitemShut {NoStop}%
\bibitem [{\citenamefont {Ralph}\ \emph {et~al.}(2007)\citenamefont {Ralph},
  \citenamefont {Resch},\ and\ \citenamefont {Gilchrist}}]{qudit_toffoli}%
  \BibitemOpen
  \bibfield  {author} {\bibinfo {author} {\bibfnamefont {T.~C.}\ \bibnamefont
  {Ralph}}, \bibinfo {author} {\bibfnamefont {K.~J.}\ \bibnamefont {Resch}},\
  and\ \bibinfo {author} {\bibfnamefont {A.}~\bibnamefont {Gilchrist}},\
  }\bibfield  {title} {\bibinfo {title} {Efficient toffoli gates using
  qudits},\ }\href {https://doi.org/10.1103/PhysRevA.75.022313} {\bibfield
  {journal} {\bibinfo  {journal} {Phys. Rev. A}\ }\textbf {\bibinfo {volume}
  {75}},\ \bibinfo {pages} {022313} (\bibinfo {year} {2007})}\BibitemShut
  {NoStop}%
\bibitem [{\citenamefont {Rico}\ \emph {et~al.}(2018)\citenamefont {Rico},
  \citenamefont {Dalmonte}, \citenamefont {Zoller}, \citenamefont {Banerjee},
  \citenamefont {B{\"o}gli}, \citenamefont {Stebler},\ and\ \citenamefont
  {Wiese}}]{SO(3)_simulation}%
  \BibitemOpen
  \bibfield  {author} {\bibinfo {author} {\bibfnamefont {E.}~\bibnamefont
  {Rico}}, \bibinfo {author} {\bibfnamefont {M.}~\bibnamefont {Dalmonte}},
  \bibinfo {author} {\bibfnamefont {P.}~\bibnamefont {Zoller}}, \bibinfo
  {author} {\bibfnamefont {D.}~\bibnamefont {Banerjee}}, \bibinfo {author}
  {\bibfnamefont {M.}~\bibnamefont {B{\"o}gli}}, \bibinfo {author}
  {\bibfnamefont {P.}~\bibnamefont {Stebler}},\ and\ \bibinfo {author}
  {\bibfnamefont {U.-J.}\ \bibnamefont {Wiese}},\ }\bibfield  {title} {\bibinfo
  {title} {So (3)“nuclear physics” with ultracold gases},\ }\href@noop {}
  {\bibfield  {journal} {\bibinfo  {journal} {Annals of physics}\ }\textbf
  {\bibinfo {volume} {393}},\ \bibinfo {pages} {466} (\bibinfo {year}
  {2018})}\BibitemShut {NoStop}%
\bibitem [{\citenamefont {Gonz\'alez-Cuadra}\ \emph {et~al.}(2022)\citenamefont
  {Gonz\'alez-Cuadra}, \citenamefont {Zache}, \citenamefont {Carrasco},
  \citenamefont {Kraus},\ and\ \citenamefont
  {Zoller}}]{qudit_simulation_rydberg}%
  \BibitemOpen
  \bibfield  {author} {\bibinfo {author} {\bibfnamefont {D.}~\bibnamefont
  {Gonz\'alez-Cuadra}}, \bibinfo {author} {\bibfnamefont {T.~V.}\ \bibnamefont
  {Zache}}, \bibinfo {author} {\bibfnamefont {J.}~\bibnamefont {Carrasco}},
  \bibinfo {author} {\bibfnamefont {B.}~\bibnamefont {Kraus}},\ and\ \bibinfo
  {author} {\bibfnamefont {P.}~\bibnamefont {Zoller}},\ }\bibfield  {title}
  {\bibinfo {title} {Hardware efficient quantum simulation of non-abelian gauge
  theories with qudits on rydberg platforms},\ }\href
  {https://doi.org/10.1103/PhysRevLett.129.160501} {\bibfield  {journal}
  {\bibinfo  {journal} {Phys. Rev. Lett.}\ }\textbf {\bibinfo {volume} {129}},\
  \bibinfo {pages} {160501} (\bibinfo {year} {2022})}\BibitemShut {NoStop}%
\bibitem [{\citenamefont {MacDonell}\ \emph {et~al.}(2021)\citenamefont
  {MacDonell}, \citenamefont {Dickerson}, \citenamefont {Birch}, \citenamefont
  {Kumar}, \citenamefont {Edmunds}, \citenamefont {Biercuk}, \citenamefont
  {Hempel},\ and\ \citenamefont {Kassal}}]{MQB_chemical_dynamics}%
  \BibitemOpen
  \bibfield  {author} {\bibinfo {author} {\bibfnamefont {R.~J.}\ \bibnamefont
  {MacDonell}}, \bibinfo {author} {\bibfnamefont {C.~E.}\ \bibnamefont
  {Dickerson}}, \bibinfo {author} {\bibfnamefont {C.~J.}\ \bibnamefont
  {Birch}}, \bibinfo {author} {\bibfnamefont {A.}~\bibnamefont {Kumar}},
  \bibinfo {author} {\bibfnamefont {C.~L.}\ \bibnamefont {Edmunds}}, \bibinfo
  {author} {\bibfnamefont {M.~J.}\ \bibnamefont {Biercuk}}, \bibinfo {author}
  {\bibfnamefont {C.}~\bibnamefont {Hempel}},\ and\ \bibinfo {author}
  {\bibfnamefont {I.}~\bibnamefont {Kassal}},\ }\bibfield  {title} {\bibinfo
  {title} {Analog quantum simulation of chemical dynamics},\ }\href@noop {}
  {\bibfield  {journal} {\bibinfo  {journal} {Chemical Science}\ }\textbf
  {\bibinfo {volume} {12}},\ \bibinfo {pages} {9794} (\bibinfo {year}
  {2021})}\BibitemShut {NoStop}%
\bibitem [{\citenamefont {Kraft}\ \emph {et~al.}(2018)\citenamefont {Kraft},
  \citenamefont {Ritz}, \citenamefont {Brunner}, \citenamefont {Huber},\ and\
  \citenamefont {G{\"u}hne}}]{multilevel_entanglement}%
  \BibitemOpen
  \bibfield  {author} {\bibinfo {author} {\bibfnamefont {T.}~\bibnamefont
  {Kraft}}, \bibinfo {author} {\bibfnamefont {C.}~\bibnamefont {Ritz}},
  \bibinfo {author} {\bibfnamefont {N.}~\bibnamefont {Brunner}}, \bibinfo
  {author} {\bibfnamefont {M.}~\bibnamefont {Huber}},\ and\ \bibinfo {author}
  {\bibfnamefont {O.}~\bibnamefont {G{\"u}hne}},\ }\bibfield  {title} {\bibinfo
  {title} {Characterizing genuine multilevel entanglement},\ }\href@noop {}
  {\bibfield  {journal} {\bibinfo  {journal} {Physical review letters}\
  }\textbf {\bibinfo {volume} {120}},\ \bibinfo {pages} {060502} (\bibinfo
  {year} {2018})}\BibitemShut {NoStop}%
\bibitem [{\citenamefont {Schaeff}\ \emph {et~al.}(2015)\citenamefont
  {Schaeff}, \citenamefont {Polster}, \citenamefont {Huber}, \citenamefont
  {Ramelow},\ and\ \citenamefont {Zeilinger}}]{intergrated_optics_entangled}%
  \BibitemOpen
  \bibfield  {author} {\bibinfo {author} {\bibfnamefont {C.}~\bibnamefont
  {Schaeff}}, \bibinfo {author} {\bibfnamefont {R.}~\bibnamefont {Polster}},
  \bibinfo {author} {\bibfnamefont {M.}~\bibnamefont {Huber}}, \bibinfo
  {author} {\bibfnamefont {S.}~\bibnamefont {Ramelow}},\ and\ \bibinfo {author}
  {\bibfnamefont {A.}~\bibnamefont {Zeilinger}},\ }\bibfield  {title} {\bibinfo
  {title} {Experimental access to higher-dimensional entangled quantum systems
  using integrated optics},\ }\href {https://doi.org/10.1364/OPTICA.2.000523}
  {\bibfield  {journal} {\bibinfo  {journal} {Optica}\ }\textbf {\bibinfo
  {volume} {2}},\ \bibinfo {pages} {523} (\bibinfo {year} {2015})}\BibitemShut
  {NoStop}%
\bibitem [{\citenamefont {Ren}\ \emph {et~al.}(2006)\citenamefont {Ren},
  \citenamefont {Guo}, \citenamefont {Huang}, \citenamefont {Li},\ and\
  \citenamefont {Guo}}]{OAM_entangled}%
  \BibitemOpen
  \bibfield  {author} {\bibinfo {author} {\bibfnamefont {X.~F.}\ \bibnamefont
  {Ren}}, \bibinfo {author} {\bibfnamefont {G.~P.}\ \bibnamefont {Guo}},
  \bibinfo {author} {\bibfnamefont {Y.~F.}\ \bibnamefont {Huang}}, \bibinfo
  {author} {\bibfnamefont {C.~F.}\ \bibnamefont {Li}},\ and\ \bibinfo {author}
  {\bibfnamefont {G.~C.}\ \bibnamefont {Guo}},\ }\bibfield  {title} {\bibinfo
  {title} {Plasmon-assisted transmission of high-dimensional orbital
  angular-momentum entangled state},\ }\href
  {https://doi.org/10.1209/epl/i2006-10359-2} {\bibfield  {journal} {\bibinfo
  {journal} {Europhysics Letters ({EPL})}\ }\textbf {\bibinfo {volume} {76}},\
  \bibinfo {pages} {753} (\bibinfo {year} {2006})}\BibitemShut {NoStop}%
\bibitem [{\citenamefont {Kues}\ \emph {et~al.}(2017)\citenamefont {Kues},
  \citenamefont {Reimer}, \citenamefont {Roztocki}, \citenamefont {Cort{\'e}s},
  \citenamefont {Sciara}, \citenamefont {Wetzel}, \citenamefont {Zhang},
  \citenamefont {Cino}, \citenamefont {Chu}, \citenamefont {Little} \emph
  {et~al.}}]{high_entangled_chip}%
  \BibitemOpen
  \bibfield  {author} {\bibinfo {author} {\bibfnamefont {M.}~\bibnamefont
  {Kues}}, \bibinfo {author} {\bibfnamefont {C.}~\bibnamefont {Reimer}},
  \bibinfo {author} {\bibfnamefont {P.}~\bibnamefont {Roztocki}}, \bibinfo
  {author} {\bibfnamefont {L.~R.}\ \bibnamefont {Cort{\'e}s}}, \bibinfo
  {author} {\bibfnamefont {S.}~\bibnamefont {Sciara}}, \bibinfo {author}
  {\bibfnamefont {B.}~\bibnamefont {Wetzel}}, \bibinfo {author} {\bibfnamefont
  {Y.}~\bibnamefont {Zhang}}, \bibinfo {author} {\bibfnamefont
  {A.}~\bibnamefont {Cino}}, \bibinfo {author} {\bibfnamefont {S.~T.}\
  \bibnamefont {Chu}}, \bibinfo {author} {\bibfnamefont {B.~E.}\ \bibnamefont
  {Little}}, \emph {et~al.},\ }\bibfield  {title} {\bibinfo {title} {On-chip
  generation of high-dimensional entangled quantum states and their coherent
  control},\ }\href@noop {} {\bibfield  {journal} {\bibinfo  {journal}
  {Nature}\ }\textbf {\bibinfo {volume} {546}},\ \bibinfo {pages} {622}
  (\bibinfo {year} {2017})}\BibitemShut {NoStop}%
\bibitem [{\citenamefont {Cervera-Lierta}\ \emph {et~al.}(2022)\citenamefont
  {Cervera-Lierta}, \citenamefont {Krenn}, \citenamefont {Aspuru-Guzik},\ and\
  \citenamefont {Galda}}]{GHZ_transmon}%
  \BibitemOpen
  \bibfield  {author} {\bibinfo {author} {\bibfnamefont {A.}~\bibnamefont
  {Cervera-Lierta}}, \bibinfo {author} {\bibfnamefont {M.}~\bibnamefont
  {Krenn}}, \bibinfo {author} {\bibfnamefont {A.}~\bibnamefont
  {Aspuru-Guzik}},\ and\ \bibinfo {author} {\bibfnamefont {A.}~\bibnamefont
  {Galda}},\ }\bibfield  {title} {\bibinfo {title} {Experimental
  high-dimensional greenberger-horne-zeilinger entanglement with
  superconducting transmon qutrits},\ }\href
  {https://doi.org/10.1103/PhysRevApplied.17.024062} {\bibfield  {journal}
  {\bibinfo  {journal} {Phys. Rev. Applied}\ }\textbf {\bibinfo {volume}
  {17}},\ \bibinfo {pages} {024062} (\bibinfo {year} {2022})}\BibitemShut
  {NoStop}%
\bibitem [{\citenamefont {Su}\ \emph {et~al.}(2022)\citenamefont {Su},
  \citenamefont {Zhang}, \citenamefont {Bin},\ and\ \citenamefont
  {Yang}}]{superconducting_qutrit}%
  \BibitemOpen
  \bibfield  {author} {\bibinfo {author} {\bibfnamefont {Q.-P.}\ \bibnamefont
  {Su}}, \bibinfo {author} {\bibfnamefont {Y.}~\bibnamefont {Zhang}}, \bibinfo
  {author} {\bibfnamefont {L.}~\bibnamefont {Bin}},\ and\ \bibinfo {author}
  {\bibfnamefont {C.-P.}\ \bibnamefont {Yang}},\ }\bibfield  {title} {\bibinfo
  {title} {Hybrid controlled-sum gate with one superconducting qutrit and one
  cat-state qutrit and application in hybrid entangled state preparation},\
  }\href {https://doi.org/10.1103/PhysRevA.105.042434} {\bibfield  {journal}
  {\bibinfo  {journal} {Phys. Rev. A}\ }\textbf {\bibinfo {volume} {105}},\
  \bibinfo {pages} {042434} (\bibinfo {year} {2022})}\BibitemShut {NoStop}%
\bibitem [{\citenamefont {Collins}\ \emph {et~al.}(2002)\citenamefont
  {Collins}, \citenamefont {Gisin}, \citenamefont {Linden}, \citenamefont
  {Massar},\ and\ \citenamefont {Popescu}}]{qudit_bell}%
  \BibitemOpen
  \bibfield  {author} {\bibinfo {author} {\bibfnamefont {D.}~\bibnamefont
  {Collins}}, \bibinfo {author} {\bibfnamefont {N.}~\bibnamefont {Gisin}},
  \bibinfo {author} {\bibfnamefont {N.}~\bibnamefont {Linden}}, \bibinfo
  {author} {\bibfnamefont {S.}~\bibnamefont {Massar}},\ and\ \bibinfo {author}
  {\bibfnamefont {S.}~\bibnamefont {Popescu}},\ }\bibfield  {title} {\bibinfo
  {title} {Bell inequalities for arbitrarily high-dimensional systems},\ }\href
  {https://doi.org/10.1103/PhysRevLett.88.040404} {\bibfield  {journal}
  {\bibinfo  {journal} {Phys. Rev. Lett.}\ }\textbf {\bibinfo {volume} {88}},\
  \bibinfo {pages} {040404} (\bibinfo {year} {2002})}\BibitemShut {NoStop}%
\bibitem [{\citenamefont {Feng}\ \emph {et~al.}(2022)\citenamefont {Feng},
  \citenamefont {Xu}, \citenamefont {Zhou}, \citenamefont {Luo}, \citenamefont
  {Zhang},\ and\ \citenamefont {Zhou}}]{212}%
  \BibitemOpen
  \bibfield  {author} {\bibinfo {author} {\bibfnamefont {T.}~\bibnamefont
  {Feng}}, \bibinfo {author} {\bibfnamefont {Q.}~\bibnamefont {Xu}}, \bibinfo
  {author} {\bibfnamefont {L.}~\bibnamefont {Zhou}}, \bibinfo {author}
  {\bibfnamefont {M.}~\bibnamefont {Luo}}, \bibinfo {author} {\bibfnamefont
  {W.}~\bibnamefont {Zhang}},\ and\ \bibinfo {author} {\bibfnamefont
  {X.}~\bibnamefont {Zhou}},\ }\bibfield  {title} {\bibinfo {title} {Quantum
  information transfer between a two-level and a four-level quantum systems},\
  }\href {https://doi.org/10.1364/PRJ.461283} {\bibfield  {journal} {\bibinfo
  {journal} {Photon. Res.}\ }\textbf {\bibinfo {volume} {10}},\ \bibinfo
  {pages} {2854} (\bibinfo {year} {2022})}\BibitemShut {NoStop}%
\bibitem [{\citenamefont {Cirac}\ \emph {et~al.}(1997)\citenamefont {Cirac},
  \citenamefont {Zoller}, \citenamefont {Kimble},\ and\ \citenamefont
  {Mabuchi}}]{PRL_state_transfer}%
  \BibitemOpen
  \bibfield  {author} {\bibinfo {author} {\bibfnamefont {J.~I.}\ \bibnamefont
  {Cirac}}, \bibinfo {author} {\bibfnamefont {P.}~\bibnamefont {Zoller}},
  \bibinfo {author} {\bibfnamefont {H.~J.}\ \bibnamefont {Kimble}},\ and\
  \bibinfo {author} {\bibfnamefont {H.}~\bibnamefont {Mabuchi}},\ }\bibfield
  {title} {\bibinfo {title} {Quantum state transfer and entanglement
  distribution among distant nodes in a quantum network},\ }\href
  {https://doi.org/10.1103/PhysRevLett.78.3221} {\bibfield  {journal} {\bibinfo
   {journal} {Phys. Rev. Lett.}\ }\textbf {\bibinfo {volume} {78}},\ \bibinfo
  {pages} {3221} (\bibinfo {year} {1997})}\BibitemShut {NoStop}%
\bibitem [{\citenamefont {Kurpiers}\ \emph {et~al.}(2018)\citenamefont
  {Kurpiers}, \citenamefont {Magnard}, \citenamefont {Walter}, \citenamefont
  {Royer}, \citenamefont {Pechal}, \citenamefont {Heinsoo}, \citenamefont
  {Salath{\'e}}, \citenamefont {Akin}, \citenamefont {Storz}, \citenamefont
  {Besse} \emph {et~al.}}]{remote-state-transfer_microwave-photons}%
  \BibitemOpen
  \bibfield  {author} {\bibinfo {author} {\bibfnamefont {P.}~\bibnamefont
  {Kurpiers}}, \bibinfo {author} {\bibfnamefont {P.}~\bibnamefont {Magnard}},
  \bibinfo {author} {\bibfnamefont {T.}~\bibnamefont {Walter}}, \bibinfo
  {author} {\bibfnamefont {B.}~\bibnamefont {Royer}}, \bibinfo {author}
  {\bibfnamefont {M.}~\bibnamefont {Pechal}}, \bibinfo {author} {\bibfnamefont
  {J.}~\bibnamefont {Heinsoo}}, \bibinfo {author} {\bibfnamefont
  {Y.}~\bibnamefont {Salath{\'e}}}, \bibinfo {author} {\bibfnamefont
  {A.}~\bibnamefont {Akin}}, \bibinfo {author} {\bibfnamefont {S.}~\bibnamefont
  {Storz}}, \bibinfo {author} {\bibfnamefont {J.-C.}\ \bibnamefont {Besse}},
  \emph {et~al.},\ }\bibfield  {title} {\bibinfo {title} {Deterministic quantum
  state transfer and remote entanglement using microwave photons},\ }\href@noop
  {} {\bibfield  {journal} {\bibinfo  {journal} {Nature}\ }\textbf {\bibinfo
  {volume} {558}},\ \bibinfo {pages} {264} (\bibinfo {year}
  {2018})}\BibitemShut {NoStop}%
\bibitem [{\citenamefont {Sillanp{\"a}{\"a}}\ \emph {et~al.}(2007)\citenamefont
  {Sillanp{\"a}{\"a}}, \citenamefont {Park},\ and\ \citenamefont
  {Simmonds}}]{state_transfer_resonant_cavity}%
  \BibitemOpen
  \bibfield  {author} {\bibinfo {author} {\bibfnamefont {M.~A.}\ \bibnamefont
  {Sillanp{\"a}{\"a}}}, \bibinfo {author} {\bibfnamefont {J.~I.}\ \bibnamefont
  {Park}},\ and\ \bibinfo {author} {\bibfnamefont {R.~W.}\ \bibnamefont
  {Simmonds}},\ }\bibfield  {title} {\bibinfo {title} {Coherent quantum state
  storage and transfer between two phase qubits via a resonant cavity},\
  }\href@noop {} {\bibfield  {journal} {\bibinfo  {journal} {Nature}\ }\textbf
  {\bibinfo {volume} {449}},\ \bibinfo {pages} {438} (\bibinfo {year}
  {2007})}\BibitemShut {NoStop}%
\bibitem [{\citenamefont {Liu}\ \emph {et~al.}(2017)\citenamefont {Liu},
  \citenamefont {Su}, \citenamefont {Yang}, \citenamefont {Zhang},
  \citenamefont {Xiong}, \citenamefont {Liu},\ and\ \citenamefont
  {Yang}}]{qudit_transfer_transmon}%
  \BibitemOpen
  \bibfield  {author} {\bibinfo {author} {\bibfnamefont {T.}~\bibnamefont
  {Liu}}, \bibinfo {author} {\bibfnamefont {Q.-P.}\ \bibnamefont {Su}},
  \bibinfo {author} {\bibfnamefont {J.-H.}\ \bibnamefont {Yang}}, \bibinfo
  {author} {\bibfnamefont {Y.}~\bibnamefont {Zhang}}, \bibinfo {author}
  {\bibfnamefont {S.-J.}\ \bibnamefont {Xiong}}, \bibinfo {author}
  {\bibfnamefont {J.-M.}\ \bibnamefont {Liu}},\ and\ \bibinfo {author}
  {\bibfnamefont {C.-P.}\ \bibnamefont {Yang}},\ }\bibfield  {title} {\bibinfo
  {title} {Transferring arbitrary d-dimensional quantum states of a
  superconducting transmon qudit in circuit qed},\ }\href@noop {} {\bibfield
  {journal} {\bibinfo  {journal} {Scientific reports}\ }\textbf {\bibinfo
  {volume} {7}},\ \bibinfo {pages} {1} (\bibinfo {year} {2017})}\BibitemShut
  {NoStop}%
\bibitem [{\citenamefont {Sigillito}\ \emph {et~al.}(2019)\citenamefont
  {Sigillito}, \citenamefont {Gullans}, \citenamefont {Edge}, \citenamefont
  {Borselli},\ and\ \citenamefont {Petta}}]{transfer_QDs}%
  \BibitemOpen
  \bibfield  {author} {\bibinfo {author} {\bibfnamefont {A.}~\bibnamefont
  {Sigillito}}, \bibinfo {author} {\bibfnamefont {M.}~\bibnamefont {Gullans}},
  \bibinfo {author} {\bibfnamefont {L.}~\bibnamefont {Edge}}, \bibinfo {author}
  {\bibfnamefont {M.}~\bibnamefont {Borselli}},\ and\ \bibinfo {author}
  {\bibfnamefont {J.}~\bibnamefont {Petta}},\ }\bibfield  {title} {\bibinfo
  {title} {Coherent transfer of quantum information in a silicon double quantum
  dot using resonant swap gates},\ }\href@noop {} {\bibfield  {journal}
  {\bibinfo  {journal} {npj Quantum Information}\ }\textbf {\bibinfo {volume}
  {5}},\ \bibinfo {pages} {1} (\bibinfo {year} {2019})}\BibitemShut {NoStop}%
\bibitem [{\citenamefont {Luis}\ \emph {et~al.}(2011)\citenamefont {Luis},
  \citenamefont {Repoll\'es}, \citenamefont {Mart\'{\i}nez-P\'erez},
  \citenamefont {Aguil\`a}, \citenamefont {Roubeau}, \citenamefont {Zueco},
  \citenamefont {Alonso}, \citenamefont {Evangelisti}, \citenamefont {Cam\'on},
  \citenamefont {Ses\'e}, \citenamefont {Barrios},\ and\ \citenamefont
  {Arom\'{\i}}}]{CNOT_SWAP_spins}%
  \BibitemOpen
  \bibfield  {author} {\bibinfo {author} {\bibfnamefont {F.}~\bibnamefont
  {Luis}}, \bibinfo {author} {\bibfnamefont {A.}~\bibnamefont {Repoll\'es}},
  \bibinfo {author} {\bibfnamefont {M.~J.}\ \bibnamefont
  {Mart\'{\i}nez-P\'erez}}, \bibinfo {author} {\bibfnamefont {D.}~\bibnamefont
  {Aguil\`a}}, \bibinfo {author} {\bibfnamefont {O.}~\bibnamefont {Roubeau}},
  \bibinfo {author} {\bibfnamefont {D.}~\bibnamefont {Zueco}}, \bibinfo
  {author} {\bibfnamefont {P.~J.}\ \bibnamefont {Alonso}}, \bibinfo {author}
  {\bibfnamefont {M.}~\bibnamefont {Evangelisti}}, \bibinfo {author}
  {\bibfnamefont {A.}~\bibnamefont {Cam\'on}}, \bibinfo {author} {\bibfnamefont
  {J.}~\bibnamefont {Ses\'e}}, \bibinfo {author} {\bibfnamefont {L.~A.}\
  \bibnamefont {Barrios}},\ and\ \bibinfo {author} {\bibfnamefont
  {G.}~\bibnamefont {Arom\'{\i}}},\ }\bibfield  {title} {\bibinfo {title}
  {Molecular prototypes for spin-based cnot and swap quantum gates},\ }\href
  {https://doi.org/10.1103/PhysRevLett.107.117203} {\bibfield  {journal}
  {\bibinfo  {journal} {Phys. Rev. Lett.}\ }\textbf {\bibinfo {volume} {107}},\
  \bibinfo {pages} {117203} (\bibinfo {year} {2011})}\BibitemShut {NoStop}%
\bibitem [{\citenamefont {Blais}\ \emph {et~al.}(2004)\citenamefont {Blais},
  \citenamefont {Huang}, \citenamefont {Wallraff}, \citenamefont {Girvin},\
  and\ \citenamefont {Schoelkopf}}]{superconducting_cQED}%
  \BibitemOpen
  \bibfield  {author} {\bibinfo {author} {\bibfnamefont {A.}~\bibnamefont
  {Blais}}, \bibinfo {author} {\bibfnamefont {R.-S.}\ \bibnamefont {Huang}},
  \bibinfo {author} {\bibfnamefont {A.}~\bibnamefont {Wallraff}}, \bibinfo
  {author} {\bibfnamefont {S.~M.}\ \bibnamefont {Girvin}},\ and\ \bibinfo
  {author} {\bibfnamefont {R.~J.}\ \bibnamefont {Schoelkopf}},\ }\bibfield
  {title} {\bibinfo {title} {Cavity quantum electrodynamics for superconducting
  electrical circuits: An architecture for quantum computation},\ }\href
  {https://doi.org/10.1103/PhysRevA.69.062320} {\bibfield  {journal} {\bibinfo
  {journal} {Phys. Rev. A}\ }\textbf {\bibinfo {volume} {69}},\ \bibinfo
  {pages} {062320} (\bibinfo {year} {2004})}\BibitemShut {NoStop}%
\bibitem [{\citenamefont {Blais}\ \emph {et~al.}(2021)\citenamefont {Blais},
  \citenamefont {Grimsmo}, \citenamefont {Girvin},\ and\ \citenamefont
  {Wallraff}}]{RMP_cirQED}%
  \BibitemOpen
  \bibfield  {author} {\bibinfo {author} {\bibfnamefont {A.}~\bibnamefont
  {Blais}}, \bibinfo {author} {\bibfnamefont {A.~L.}\ \bibnamefont {Grimsmo}},
  \bibinfo {author} {\bibfnamefont {S.~M.}\ \bibnamefont {Girvin}},\ and\
  \bibinfo {author} {\bibfnamefont {A.}~\bibnamefont {Wallraff}},\ }\bibfield
  {title} {\bibinfo {title} {Circuit quantum electrodynamics},\ }\href
  {https://doi.org/10.1103/RevModPhys.93.025005} {\bibfield  {journal}
  {\bibinfo  {journal} {Rev. Mod. Phys.}\ }\textbf {\bibinfo {volume} {93}},\
  \bibinfo {pages} {025005} (\bibinfo {year} {2021})}\BibitemShut {NoStop}%
\bibitem [{\citenamefont {Wineland}\ \emph {et~al.}(1998)\citenamefont
  {Wineland}, \citenamefont {Monroe}, \citenamefont {Itano}, \citenamefont
  {King}, \citenamefont {Leibfried}, \citenamefont {Myatt},\ and\ \citenamefont
  {Wood}}]{ion_simulator}%
  \BibitemOpen
  \bibfield  {author} {\bibinfo {author} {\bibfnamefont {D.~J.}\ \bibnamefont
  {Wineland}}, \bibinfo {author} {\bibfnamefont {C.}~\bibnamefont {Monroe}},
  \bibinfo {author} {\bibfnamefont {W.~M.}\ \bibnamefont {Itano}}, \bibinfo
  {author} {\bibfnamefont {B.~E.}\ \bibnamefont {King}}, \bibinfo {author}
  {\bibfnamefont {D.}~\bibnamefont {Leibfried}}, \bibinfo {author}
  {\bibfnamefont {C.}~\bibnamefont {Myatt}},\ and\ \bibinfo {author}
  {\bibfnamefont {C.}~\bibnamefont {Wood}},\ }\bibfield  {title} {\bibinfo
  {title} {Trapped-ion quantum simulator},\ }\href
  {https://doi.org/10.1238/physica.topical.076a00147} {\bibfield  {journal}
  {\bibinfo  {journal} {Physica Scripta}\ }\textbf {\bibinfo {volume} {T76}},\
  \bibinfo {pages} {147} (\bibinfo {year} {1998})}\BibitemShut {NoStop}%
\bibitem [{\citenamefont {Kasper}\ \emph {et~al.}(2017)\citenamefont {Kasper},
  \citenamefont {Hebenstreit}, \citenamefont {Jendrzejewski}, \citenamefont
  {Oberthaler},\ and\ \citenamefont {Berges}}]{ultracold_QED}%
  \BibitemOpen
  \bibfield  {author} {\bibinfo {author} {\bibfnamefont {V.}~\bibnamefont
  {Kasper}}, \bibinfo {author} {\bibfnamefont {F.}~\bibnamefont {Hebenstreit}},
  \bibinfo {author} {\bibfnamefont {F.}~\bibnamefont {Jendrzejewski}}, \bibinfo
  {author} {\bibfnamefont {M.~K.}\ \bibnamefont {Oberthaler}},\ and\ \bibinfo
  {author} {\bibfnamefont {J.}~\bibnamefont {Berges}},\ }\bibfield  {title}
  {\bibinfo {title} {Implementing quantum electrodynamics with ultracold atomic
  systems},\ }\href {https://doi.org/10.1088/1367-2630/aa54e0} {\bibfield
  {journal} {\bibinfo  {journal} {New Journal of Physics}\ }\textbf {\bibinfo
  {volume} {19}},\ \bibinfo {pages} {023030} (\bibinfo {year}
  {2017})}\BibitemShut {NoStop}%
\bibitem [{\citenamefont {Kiraz}\ \emph {et~al.}(2001)\citenamefont {Kiraz},
  \citenamefont {Michler}, \citenamefont {Becher}, \citenamefont {Gayral},
  \citenamefont {Imamo{\u{g}}lu}, \citenamefont {Zhang}, \citenamefont {Hu},
  \citenamefont {Schoenfeld},\ and\ \citenamefont {Petroff}}]{kiraz2001cavity}%
  \BibitemOpen
  \bibfield  {author} {\bibinfo {author} {\bibfnamefont {A.}~\bibnamefont
  {Kiraz}}, \bibinfo {author} {\bibfnamefont {P.}~\bibnamefont {Michler}},
  \bibinfo {author} {\bibfnamefont {C.}~\bibnamefont {Becher}}, \bibinfo
  {author} {\bibfnamefont {B.}~\bibnamefont {Gayral}}, \bibinfo {author}
  {\bibfnamefont {A.}~\bibnamefont {Imamo{\u{g}}lu}}, \bibinfo {author}
  {\bibfnamefont {L.}~\bibnamefont {Zhang}}, \bibinfo {author} {\bibfnamefont
  {E.}~\bibnamefont {Hu}}, \bibinfo {author} {\bibfnamefont {W.}~\bibnamefont
  {Schoenfeld}},\ and\ \bibinfo {author} {\bibfnamefont {P.}~\bibnamefont
  {Petroff}},\ }\bibfield  {title} {\bibinfo {title} {Cavity-quantum
  electrodynamics using a single inas quantum dot in a microdisk structure},\
  }\href@noop {} {\bibfield  {journal} {\bibinfo  {journal} {Applied Physics
  Letters}\ }\textbf {\bibinfo {volume} {78}},\ \bibinfo {pages} {3932}
  (\bibinfo {year} {2001})}\BibitemShut {NoStop}%
\bibitem [{\citenamefont {Vu{\v{c}}kovi{\'c}}\ and\ \citenamefont
  {Yamamoto}(2003)}]{vuvckovic2003photonic}%
  \BibitemOpen
  \bibfield  {author} {\bibinfo {author} {\bibfnamefont {J.}~\bibnamefont
  {Vu{\v{c}}kovi{\'c}}}\ and\ \bibinfo {author} {\bibfnamefont
  {Y.}~\bibnamefont {Yamamoto}},\ }\bibfield  {title} {\bibinfo {title}
  {Photonic crystal microcavities for cavity quantum electrodynamics with a
  single quantum dot},\ }\href@noop {} {\bibfield  {journal} {\bibinfo
  {journal} {Applied Physics Letters}\ }\textbf {\bibinfo {volume} {82}},\
  \bibinfo {pages} {2374} (\bibinfo {year} {2003})}\BibitemShut {NoStop}%
\bibitem [{\citenamefont {Domokos}\ \emph {et~al.}(1995)\citenamefont
  {Domokos}, \citenamefont {Raimond}, \citenamefont {Brune},\ and\
  \citenamefont {Haroche}}]{simple_cQED_2bit}%
  \BibitemOpen
  \bibfield  {author} {\bibinfo {author} {\bibfnamefont {P.}~\bibnamefont
  {Domokos}}, \bibinfo {author} {\bibfnamefont {J.-M.}\ \bibnamefont
  {Raimond}}, \bibinfo {author} {\bibfnamefont {M.}~\bibnamefont {Brune}},\
  and\ \bibinfo {author} {\bibfnamefont {S.}~\bibnamefont {Haroche}},\
  }\bibfield  {title} {\bibinfo {title} {Simple cavity-qed two-bit universal
  quantum logic gate: The principle and expected performances},\ }\href@noop {}
  {\bibfield  {journal} {\bibinfo  {journal} {Physical Review A}\ }\textbf
  {\bibinfo {volume} {52}},\ \bibinfo {pages} {3554} (\bibinfo {year}
  {1995})}\BibitemShut {NoStop}%
\bibitem [{\citenamefont {Ladd}\ and\ \citenamefont
  {Yamamoto}(2011)}]{QD_cQED_gate}%
  \BibitemOpen
  \bibfield  {author} {\bibinfo {author} {\bibfnamefont {T.}~\bibnamefont
  {Ladd}}\ and\ \bibinfo {author} {\bibfnamefont {Y.}~\bibnamefont
  {Yamamoto}},\ }\bibfield  {title} {\bibinfo {title} {Simple quantum logic
  gate with quantum dot cavity qed systems},\ }\href@noop {} {\bibfield
  {journal} {\bibinfo  {journal} {Physical Review B}\ }\textbf {\bibinfo
  {volume} {84}},\ \bibinfo {pages} {235307} (\bibinfo {year}
  {2011})}\BibitemShut {NoStop}%
\bibitem [{\citenamefont {Yang}\ \emph {et~al.}(2003)\citenamefont {Yang},
  \citenamefont {Chu},\ and\ \citenamefont {Han}}]{SQID_gate}%
  \BibitemOpen
  \bibfield  {author} {\bibinfo {author} {\bibfnamefont {C.-P.}\ \bibnamefont
  {Yang}}, \bibinfo {author} {\bibfnamefont {S.-I.}\ \bibnamefont {Chu}},\ and\
  \bibinfo {author} {\bibfnamefont {S.}~\bibnamefont {Han}},\ }\bibfield
  {title} {\bibinfo {title} {Possible realization of entanglement, logical
  gates, and quantum-information transfer with
  superconducting-quantum-interference-device qubits in cavity qed},\
  }\href@noop {} {\bibfield  {journal} {\bibinfo  {journal} {Physical Review
  A}\ }\textbf {\bibinfo {volume} {67}},\ \bibinfo {pages} {042311} (\bibinfo
  {year} {2003})}\BibitemShut {NoStop}%
\bibitem [{\citenamefont {Bennett}\ \emph {et~al.}(1993)\citenamefont
  {Bennett}, \citenamefont {Brassard}, \citenamefont {Cr\'epeau}, \citenamefont
  {Jozsa}, \citenamefont {Peres},\ and\ \citenamefont {Wootters}}]{PRL_tele}%
  \BibitemOpen
  \bibfield  {author} {\bibinfo {author} {\bibfnamefont {C.~H.}\ \bibnamefont
  {Bennett}}, \bibinfo {author} {\bibfnamefont {G.}~\bibnamefont {Brassard}},
  \bibinfo {author} {\bibfnamefont {C.}~\bibnamefont {Cr\'epeau}}, \bibinfo
  {author} {\bibfnamefont {R.}~\bibnamefont {Jozsa}}, \bibinfo {author}
  {\bibfnamefont {A.}~\bibnamefont {Peres}},\ and\ \bibinfo {author}
  {\bibfnamefont {W.~K.}\ \bibnamefont {Wootters}},\ }\bibfield  {title}
  {\bibinfo {title} {Teleporting an unknown quantum state via dual classical
  and einstein-podolsky-rosen channels},\ }\href
  {https://doi.org/10.1103/PhysRevLett.70.1895} {\bibfield  {journal} {\bibinfo
   {journal} {Phys. Rev. Lett.}\ }\textbf {\bibinfo {volume} {70}},\ \bibinfo
  {pages} {1895} (\bibinfo {year} {1993})}\BibitemShut {NoStop}%
\bibitem [{\citenamefont {Duan}\ \emph {et~al.}(2001)\citenamefont {Duan},
  \citenamefont {Lukin}, \citenamefont {Cirac},\ and\ \citenamefont
  {Zoller}}]{Nature_DLCZ}%
  \BibitemOpen
  \bibfield  {author} {\bibinfo {author} {\bibfnamefont {L.-M.}\ \bibnamefont
  {Duan}}, \bibinfo {author} {\bibfnamefont {M.~D.}\ \bibnamefont {Lukin}},
  \bibinfo {author} {\bibfnamefont {J.~I.}\ \bibnamefont {Cirac}},\ and\
  \bibinfo {author} {\bibfnamefont {P.}~\bibnamefont {Zoller}},\ }\bibfield
  {title} {\bibinfo {title} {Long-distance quantum communication with atomic
  ensembles and linear optics},\ }\href@noop {} {\bibfield  {journal} {\bibinfo
   {journal} {Nature}\ }\textbf {\bibinfo {volume} {414}},\ \bibinfo {pages}
  {413} (\bibinfo {year} {2001})}\BibitemShut {NoStop}%
\bibitem [{\citenamefont {Goyal}\ \emph {et~al.}(2014)\citenamefont {Goyal},
  \citenamefont {Boukama-Dzoussi}, \citenamefont {Ghosh}, \citenamefont
  {Roux},\ and\ \citenamefont {Konrad}}]{qudit_teleportation_linear-optics}%
  \BibitemOpen
  \bibfield  {author} {\bibinfo {author} {\bibfnamefont {S.~K.}\ \bibnamefont
  {Goyal}}, \bibinfo {author} {\bibfnamefont {P.~E.}\ \bibnamefont
  {Boukama-Dzoussi}}, \bibinfo {author} {\bibfnamefont {S.}~\bibnamefont
  {Ghosh}}, \bibinfo {author} {\bibfnamefont {F.~S.}\ \bibnamefont {Roux}},\
  and\ \bibinfo {author} {\bibfnamefont {T.}~\bibnamefont {Konrad}},\
  }\bibfield  {title} {\bibinfo {title} {Qudit-teleportation for photons with
  linear optics},\ }\href@noop {} {\bibfield  {journal} {\bibinfo  {journal}
  {Scientific reports}\ }\textbf {\bibinfo {volume} {4}},\ \bibinfo {pages} {1}
  (\bibinfo {year} {2014})}\BibitemShut {NoStop}%
\bibitem [{\citenamefont {Zheng}\ \emph {et~al.}(2022)\citenamefont {Zheng},
  \citenamefont {Zhang}, \citenamefont {Dong}, \citenamefont {Xu},
  \citenamefont {Wang}, \citenamefont {Wang}, \citenamefont {Li}, \citenamefont
  {Lan}, \citenamefont {Zhao}, \citenamefont {Li} \emph
  {et~al.}}]{STRAP_qudit}%
  \BibitemOpen
  \bibfield  {author} {\bibinfo {author} {\bibfnamefont {W.}~\bibnamefont
  {Zheng}}, \bibinfo {author} {\bibfnamefont {Y.}~\bibnamefont {Zhang}},
  \bibinfo {author} {\bibfnamefont {Y.}~\bibnamefont {Dong}}, \bibinfo {author}
  {\bibfnamefont {J.}~\bibnamefont {Xu}}, \bibinfo {author} {\bibfnamefont
  {Z.}~\bibnamefont {Wang}}, \bibinfo {author} {\bibfnamefont {X.}~\bibnamefont
  {Wang}}, \bibinfo {author} {\bibfnamefont {Y.}~\bibnamefont {Li}}, \bibinfo
  {author} {\bibfnamefont {D.}~\bibnamefont {Lan}}, \bibinfo {author}
  {\bibfnamefont {J.}~\bibnamefont {Zhao}}, \bibinfo {author} {\bibfnamefont
  {S.}~\bibnamefont {Li}}, \emph {et~al.},\ }\bibfield  {title} {\bibinfo
  {title} {Optimal control of stimulated raman adiabatic passage in a
  superconducting qudit},\ }\href@noop {} {\bibfield  {journal} {\bibinfo
  {journal} {npj Quantum Information}\ }\textbf {\bibinfo {volume} {8}},\
  \bibinfo {pages} {1} (\bibinfo {year} {2022})}\BibitemShut {NoStop}%
\bibitem [{\citenamefont {Zhang}\ \emph {et~al.}(2019)\citenamefont {Zhang},
  \citenamefont {Chen}, \citenamefont {Cui}, \citenamefont {Dowling},
  \citenamefont {Ou},\ and\ \citenamefont
  {Byrnes}}]{qudit_teleportation_photonic}%
  \BibitemOpen
  \bibfield  {author} {\bibinfo {author} {\bibfnamefont {C.}~\bibnamefont
  {Zhang}}, \bibinfo {author} {\bibfnamefont {J.~F.}\ \bibnamefont {Chen}},
  \bibinfo {author} {\bibfnamefont {C.}~\bibnamefont {Cui}}, \bibinfo {author}
  {\bibfnamefont {J.~P.}\ \bibnamefont {Dowling}}, \bibinfo {author}
  {\bibfnamefont {Z.~Y.}\ \bibnamefont {Ou}},\ and\ \bibinfo {author}
  {\bibfnamefont {T.}~\bibnamefont {Byrnes}},\ }\bibfield  {title} {\bibinfo
  {title} {Quantum teleportation of photonic qudits using linear optics},\
  }\href {https://doi.org/10.1103/PhysRevA.100.032330} {\bibfield  {journal}
  {\bibinfo  {journal} {Phys. Rev. A}\ }\textbf {\bibinfo {volume} {100}},\
  \bibinfo {pages} {032330} (\bibinfo {year} {2019})}\BibitemShut {NoStop}%
\bibitem [{\citenamefont {Luo}\ \emph {et~al.}(2016)\citenamefont {Luo},
  \citenamefont {Li}, \citenamefont {Lai},\ and\ \citenamefont
  {Wang}}]{teleportation_ququart}%
  \BibitemOpen
  \bibfield  {author} {\bibinfo {author} {\bibfnamefont {M.-X.}\ \bibnamefont
  {Luo}}, \bibinfo {author} {\bibfnamefont {H.-R.}\ \bibnamefont {Li}},
  \bibinfo {author} {\bibfnamefont {H.}~\bibnamefont {Lai}},\ and\ \bibinfo
  {author} {\bibfnamefont {X.}~\bibnamefont {Wang}},\ }\bibfield  {title}
  {\bibinfo {title} {Teleportation of a ququart system using hyperentangled
  photons assisted by atomic-ensemble memories},\ }\href
  {https://doi.org/10.1103/PhysRevA.93.012332} {\bibfield  {journal} {\bibinfo
  {journal} {Phys. Rev. A}\ }\textbf {\bibinfo {volume} {93}},\ \bibinfo
  {pages} {012332} (\bibinfo {year} {2016})}\BibitemShut {NoStop}%
\bibitem [{\citenamefont {Xi-Han}\ \emph {et~al.}(2007)\citenamefont {Xi-Han},
  \citenamefont {Fu-Guo},\ and\ \citenamefont
  {Hong-Yu}}]{qudit_teleportation_GHZ}%
  \BibitemOpen
  \bibfield  {author} {\bibinfo {author} {\bibfnamefont {L.}~\bibnamefont
  {Xi-Han}}, \bibinfo {author} {\bibfnamefont {D.}~\bibnamefont {Fu-Guo}},\
  and\ \bibinfo {author} {\bibfnamefont {Z.}~\bibnamefont {Hong-Yu}},\
  }\bibfield  {title} {\bibinfo {title} {Controlled teleportation of an
  arbitrary multi-qudit state in a general form with d-dimensional
  greenberger–horne–zeilinger states},\ }\href
  {https://doi.org/10.1088/0256-307X/24/5/007} {\bibfield  {journal} {\bibinfo
  {journal} {Chinese Physics Letters}\ }\textbf {\bibinfo {volume} {24}},\
  \bibinfo {pages} {1151} (\bibinfo {year} {2007})}\BibitemShut {NoStop}%
\bibitem [{\citenamefont {Zheng}\ and\ \citenamefont
  {Guo}(2000)}]{two-atom_cQED}%
  \BibitemOpen
  \bibfield  {author} {\bibinfo {author} {\bibfnamefont {S.-B.}\ \bibnamefont
  {Zheng}}\ and\ \bibinfo {author} {\bibfnamefont {G.-C.}\ \bibnamefont
  {Guo}},\ }\bibfield  {title} {\bibinfo {title} {Efficient scheme for two-atom
  entanglement and quantum information processing in cavity qed},\ }\href
  {https://doi.org/10.1103/PhysRevLett.85.2392} {\bibfield  {journal} {\bibinfo
   {journal} {Phys. Rev. Lett.}\ }\textbf {\bibinfo {volume} {85}},\ \bibinfo
  {pages} {2392} (\bibinfo {year} {2000})}\BibitemShut {NoStop}%
\bibitem [{\citenamefont {Guo}\ \emph {et~al.}(2002)\citenamefont {Guo},
  \citenamefont {Li}, \citenamefont {Li},\ and\ \citenamefont
  {Guo}}]{multiparticle_cQED}%
  \BibitemOpen
  \bibfield  {author} {\bibinfo {author} {\bibfnamefont {G.-P.}\ \bibnamefont
  {Guo}}, \bibinfo {author} {\bibfnamefont {C.-F.}\ \bibnamefont {Li}},
  \bibinfo {author} {\bibfnamefont {J.}~\bibnamefont {Li}},\ and\ \bibinfo
  {author} {\bibfnamefont {G.-C.}\ \bibnamefont {Guo}},\ }\bibfield  {title}
  {\bibinfo {title} {Scheme for the preparation of multiparticle entanglement
  in cavity qed},\ }\href {https://doi.org/10.1103/PhysRevA.65.042102}
  {\bibfield  {journal} {\bibinfo  {journal} {Phys. Rev. A}\ }\textbf {\bibinfo
  {volume} {65}},\ \bibinfo {pages} {042102} (\bibinfo {year}
  {2002})}\BibitemShut {NoStop}%
\bibitem [{\citenamefont {Guo}\ and\ \citenamefont {Zhang}(2002)}]{W_cQED}%
  \BibitemOpen
  \bibfield  {author} {\bibinfo {author} {\bibfnamefont {G.-C.}\ \bibnamefont
  {Guo}}\ and\ \bibinfo {author} {\bibfnamefont {Y.-S.}\ \bibnamefont
  {Zhang}},\ }\bibfield  {title} {\bibinfo {title} {Scheme for preparation of
  the w state via cavity quantum electrodynamics},\ }\href
  {https://doi.org/10.1103/PhysRevA.65.054302} {\bibfield  {journal} {\bibinfo
  {journal} {Phys. Rev. A}\ }\textbf {\bibinfo {volume} {65}},\ \bibinfo
  {pages} {054302} (\bibinfo {year} {2002})}\BibitemShut {NoStop}%
\bibitem [{\citenamefont {Zhou}\ \emph {et~al.}(2022)\citenamefont {Zhou},
  \citenamefont {Xu}, \citenamefont {Feng},\ and\ \citenamefont
  {Zhou}}]{224_photon}%
  \BibitemOpen
  \bibfield  {author} {\bibinfo {author} {\bibfnamefont {L.}~\bibnamefont
  {Zhou}}, \bibinfo {author} {\bibfnamefont {Q.}~\bibnamefont {Xu}}, \bibinfo
  {author} {\bibfnamefont {T.}~\bibnamefont {Feng}},\ and\ \bibinfo {author}
  {\bibfnamefont {X.}~\bibnamefont {Zhou}},\ }\href@noop {} {\bibinfo {title}
  {Experimental realization of a three-photon asymmetric maximally entangled
  state and its application to quantum teleportation}} (\bibinfo {year}
  {2022}),\ \Eprint {https://arxiv.org/abs/2212.00545} {arXiv:2212.00545
  [quant-ph]} \BibitemShut {NoStop}%
\bibitem [{\citenamefont {Schrieffer}\ and\ \citenamefont
  {Wolff}(1966)}]{SWtransformation}%
  \BibitemOpen
  \bibfield  {author} {\bibinfo {author} {\bibfnamefont {J.~R.}\ \bibnamefont
  {Schrieffer}}\ and\ \bibinfo {author} {\bibfnamefont {P.~A.}\ \bibnamefont
  {Wolff}},\ }\bibfield  {title} {\bibinfo {title} {Relation between the
  anderson and kondo hamiltonians},\ }\href
  {https://doi.org/10.1103/PhysRev.149.491} {\bibfield  {journal} {\bibinfo
  {journal} {Phys. Rev.}\ }\textbf {\bibinfo {volume} {149}},\ \bibinfo {pages}
  {491} (\bibinfo {year} {1966})}\BibitemShut {NoStop}%
\bibitem [{\citenamefont {Tang}\ \emph {et~al.}(2019)\citenamefont {Tang},
  \citenamefont {Tang}, \citenamefont {Zhang}, \citenamefont {Lu},
  \citenamefont {Zhang}, \citenamefont {Zhang}, \citenamefont {Xia},\ and\
  \citenamefont {Xiao}}]{chip_chiral-interface}%
  \BibitemOpen
  \bibfield  {author} {\bibinfo {author} {\bibfnamefont {L.}~\bibnamefont
  {Tang}}, \bibinfo {author} {\bibfnamefont {J.}~\bibnamefont {Tang}}, \bibinfo
  {author} {\bibfnamefont {W.}~\bibnamefont {Zhang}}, \bibinfo {author}
  {\bibfnamefont {G.}~\bibnamefont {Lu}}, \bibinfo {author} {\bibfnamefont
  {H.}~\bibnamefont {Zhang}}, \bibinfo {author} {\bibfnamefont
  {Y.}~\bibnamefont {Zhang}}, \bibinfo {author} {\bibfnamefont
  {K.}~\bibnamefont {Xia}},\ and\ \bibinfo {author} {\bibfnamefont
  {M.}~\bibnamefont {Xiao}},\ }\bibfield  {title} {\bibinfo {title} {On-chip
  chiral single-photon interface: Isolation and unidirectional emission},\
  }\href {https://doi.org/10.1103/PhysRevA.99.043833} {\bibfield  {journal}
  {\bibinfo  {journal} {Phys. Rev. A}\ }\textbf {\bibinfo {volume} {99}},\
  \bibinfo {pages} {043833} (\bibinfo {year} {2019})}\BibitemShut {NoStop}%
\bibitem [{\citenamefont {Sayrin}\ \emph {et~al.}(2015)\citenamefont {Sayrin},
  \citenamefont {Junge}, \citenamefont {Mitsch}, \citenamefont {Albrecht},
  \citenamefont {O'Shea}, \citenamefont {Schneeweiss}, \citenamefont {Volz},\
  and\ \citenamefont {Rauschenbeutel}}]{isolator_cold-atoms}%
  \BibitemOpen
  \bibfield  {author} {\bibinfo {author} {\bibfnamefont {C.}~\bibnamefont
  {Sayrin}}, \bibinfo {author} {\bibfnamefont {C.}~\bibnamefont {Junge}},
  \bibinfo {author} {\bibfnamefont {R.}~\bibnamefont {Mitsch}}, \bibinfo
  {author} {\bibfnamefont {B.}~\bibnamefont {Albrecht}}, \bibinfo {author}
  {\bibfnamefont {D.}~\bibnamefont {O'Shea}}, \bibinfo {author} {\bibfnamefont
  {P.}~\bibnamefont {Schneeweiss}}, \bibinfo {author} {\bibfnamefont
  {J.}~\bibnamefont {Volz}},\ and\ \bibinfo {author} {\bibfnamefont
  {A.}~\bibnamefont {Rauschenbeutel}},\ }\bibfield  {title} {\bibinfo {title}
  {Nanophotonic optical isolator controlled by the internal state of cold
  atoms},\ }\href {https://doi.org/10.1103/PhysRevX.5.041036} {\bibfield
  {journal} {\bibinfo  {journal} {Phys. Rev. X}\ }\textbf {\bibinfo {volume}
  {5}},\ \bibinfo {pages} {041036} (\bibinfo {year} {2015})}\BibitemShut
  {NoStop}%
\bibitem [{\citenamefont {Ren}\ \emph {et~al.}(2022)\citenamefont {Ren},
  \citenamefont {Franke},\ and\ \citenamefont
  {Hughes}}]{microring-resonators_linearly-polarized}%
  \BibitemOpen
  \bibfield  {author} {\bibinfo {author} {\bibfnamefont {J.}~\bibnamefont
  {Ren}}, \bibinfo {author} {\bibfnamefont {S.}~\bibnamefont {Franke}},\ and\
  \bibinfo {author} {\bibfnamefont {S.}~\bibnamefont {Hughes}},\ }\bibfield
  {title} {\bibinfo {title} {Quasinormal mode theory of chiral power flow from
  linearly polarized dipole emitters coupled to index-modulated microring
  resonators close to an exceptional point},\ }\href@noop {} {\bibfield
  {journal} {\bibinfo  {journal} {ACS Photonics}\ }\textbf {\bibinfo {volume}
  {9}},\ \bibinfo {pages} {1315} (\bibinfo {year} {2022})}\BibitemShut
  {NoStop}%
\bibitem [{\citenamefont {Scheucher}\ \emph {et~al.}(2016)\citenamefont
  {Scheucher}, \citenamefont {Hilico}, \citenamefont {Will}, \citenamefont
  {Volz},\ and\ \citenamefont {Rauschenbeutel}}]{optical_circulator}%
  \BibitemOpen
  \bibfield  {author} {\bibinfo {author} {\bibfnamefont {M.}~\bibnamefont
  {Scheucher}}, \bibinfo {author} {\bibfnamefont {A.}~\bibnamefont {Hilico}},
  \bibinfo {author} {\bibfnamefont {E.}~\bibnamefont {Will}}, \bibinfo {author}
  {\bibfnamefont {J.}~\bibnamefont {Volz}},\ and\ \bibinfo {author}
  {\bibfnamefont {A.}~\bibnamefont {Rauschenbeutel}},\ }\bibfield  {title}
  {\bibinfo {title} {Quantum optical circulator controlled by a single chirally
  coupled atom},\ }\href {https://doi.org/10.1126/science.aaj2118} {\bibfield
  {journal} {\bibinfo  {journal} {Science}\ }\textbf {\bibinfo {volume}
  {354}},\ \bibinfo {pages} {1577} (\bibinfo {year} {2016})},\ \Eprint
  {https://arxiv.org/abs/https://www.science.org/doi/pdf/10.1126/science.aaj2118}
  {https://www.science.org/doi/pdf/10.1126/science.aaj2118} \BibitemShut
  {NoStop}%
\bibitem [{\citenamefont {Peter}\ \emph {et~al.}(2005)\citenamefont {Peter},
  \citenamefont {Senellart}, \citenamefont {Martrou}, \citenamefont
  {Lema\^{\i}tre}, \citenamefont {Hours}, \citenamefont {G\'erard},\ and\
  \citenamefont {Bloch}}]{Exciton-Photon_strong_coupling_microcavity}%
  \BibitemOpen
  \bibfield  {author} {\bibinfo {author} {\bibfnamefont {E.}~\bibnamefont
  {Peter}}, \bibinfo {author} {\bibfnamefont {P.}~\bibnamefont {Senellart}},
  \bibinfo {author} {\bibfnamefont {D.}~\bibnamefont {Martrou}}, \bibinfo
  {author} {\bibfnamefont {A.}~\bibnamefont {Lema\^{\i}tre}}, \bibinfo {author}
  {\bibfnamefont {J.}~\bibnamefont {Hours}}, \bibinfo {author} {\bibfnamefont
  {J.~M.}\ \bibnamefont {G\'erard}},\ and\ \bibinfo {author} {\bibfnamefont
  {J.}~\bibnamefont {Bloch}},\ }\bibfield  {title} {\bibinfo {title}
  {Exciton-photon strong-coupling regime for a single quantum dot embedded in a
  microcavity},\ }\href {https://doi.org/10.1103/PhysRevLett.95.067401}
  {\bibfield  {journal} {\bibinfo  {journal} {Phys. Rev. Lett.}\ }\textbf
  {\bibinfo {volume} {95}},\ \bibinfo {pages} {067401} (\bibinfo {year}
  {2005})}\BibitemShut {NoStop}%
\bibitem [{\citenamefont {Lodahl}\ \emph {et~al.}(2015)\citenamefont {Lodahl},
  \citenamefont {Mahmoodian},\ and\ \citenamefont
  {Stobbe}}]{rmp_interfacing_qd}%
  \BibitemOpen
  \bibfield  {author} {\bibinfo {author} {\bibfnamefont {P.}~\bibnamefont
  {Lodahl}}, \bibinfo {author} {\bibfnamefont {S.}~\bibnamefont {Mahmoodian}},\
  and\ \bibinfo {author} {\bibfnamefont {S.}~\bibnamefont {Stobbe}},\
  }\bibfield  {title} {\bibinfo {title} {Interfacing single photons and single
  quantum dots with photonic nanostructures},\ }\href
  {https://doi.org/10.1103/RevModPhys.87.347} {\bibfield  {journal} {\bibinfo
  {journal} {Rev. Mod. Phys.}\ }\textbf {\bibinfo {volume} {87}},\ \bibinfo
  {pages} {347} (\bibinfo {year} {2015})}\BibitemShut {NoStop}%
\bibitem [{\citenamefont {Takagahara}(1989)}]{QD_biexciton}%
  \BibitemOpen
  \bibfield  {author} {\bibinfo {author} {\bibfnamefont {T.}~\bibnamefont
  {Takagahara}},\ }\bibfield  {title} {\bibinfo {title} {Biexciton states in
  semiconductor quantum dots and their nonlinear optical properties},\ }\href
  {https://doi.org/10.1103/PhysRevB.39.10206} {\bibfield  {journal} {\bibinfo
  {journal} {Phys. Rev. B}\ }\textbf {\bibinfo {volume} {39}},\ \bibinfo
  {pages} {10206} (\bibinfo {year} {1989})}\BibitemShut {NoStop}%
\bibitem [{\citenamefont {Poshakinskiy}\ and\ \citenamefont
  {Poddubny}(2016)}]{QD_biexciton_blockade}%
  \BibitemOpen
  \bibfield  {author} {\bibinfo {author} {\bibfnamefont {A.~V.}\ \bibnamefont
  {Poshakinskiy}}\ and\ \bibinfo {author} {\bibfnamefont {A.~N.}\ \bibnamefont
  {Poddubny}},\ }\bibfield  {title} {\bibinfo {title} {Biexciton-mediated
  superradiant photon blockade},\ }\href
  {https://doi.org/10.1103/PhysRevA.93.033856} {\bibfield  {journal} {\bibinfo
  {journal} {Phys. Rev. A}\ }\textbf {\bibinfo {volume} {93}},\ \bibinfo
  {pages} {033856} (\bibinfo {year} {2016})}\BibitemShut {NoStop}%
\bibitem [{\citenamefont {Xu}\ \emph {et~al.}(2008)\citenamefont {Xu},
  \citenamefont {Fattal},\ and\ \citenamefont
  {Beausoleil}}]{silicon_microring}%
  \BibitemOpen
  \bibfield  {author} {\bibinfo {author} {\bibfnamefont {Q.}~\bibnamefont
  {Xu}}, \bibinfo {author} {\bibfnamefont {D.}~\bibnamefont {Fattal}},\ and\
  \bibinfo {author} {\bibfnamefont {R.~G.}\ \bibnamefont {Beausoleil}},\
  }\bibfield  {title} {\bibinfo {title} {Silicon microring resonators with
  1.5-{\textmu}m radius},\ }\href {https://doi.org/10.1364/OE.16.004309}
  {\bibfield  {journal} {\bibinfo  {journal} {Opt. Express}\ }\textbf {\bibinfo
  {volume} {16}},\ \bibinfo {pages} {4309} (\bibinfo {year}
  {2008})}\BibitemShut {NoStop}%
\bibitem [{\citenamefont {He}\ \emph {et~al.}(2010)\citenamefont {He},
  \citenamefont {Yang}, \citenamefont {Li}, \citenamefont {Luo},\ and\
  \citenamefont {Han}}]{4-level_SQUID}%
  \BibitemOpen
  \bibfield  {author} {\bibinfo {author} {\bibfnamefont {X.-L.}\ \bibnamefont
  {He}}, \bibinfo {author} {\bibfnamefont {C.-P.}\ \bibnamefont {Yang}},
  \bibinfo {author} {\bibfnamefont {S.}~\bibnamefont {Li}}, \bibinfo {author}
  {\bibfnamefont {J.-Y.}\ \bibnamefont {Luo}},\ and\ \bibinfo {author}
  {\bibfnamefont {S.}~\bibnamefont {Han}},\ }\bibfield  {title} {\bibinfo
  {title} {Quantum logical gates with four-level superconducting quantum
  interference devices coupled to a superconducting resonator},\ }\href
  {https://doi.org/10.1103/PhysRevA.82.024301} {\bibfield  {journal} {\bibinfo
  {journal} {Phys. Rev. A}\ }\textbf {\bibinfo {volume} {82}},\ \bibinfo
  {pages} {024301} (\bibinfo {year} {2010})}\BibitemShut {NoStop}%
\bibitem [{\citenamefont {Waseem}\ \emph {et~al.}(2012)\citenamefont {Waseem},
  \citenamefont {Irfan},\ and\ \citenamefont {Qamar}}]{4-level_SQUID-SR}%
  \BibitemOpen
  \bibfield  {author} {\bibinfo {author} {\bibfnamefont {M.}~\bibnamefont
  {Waseem}}, \bibinfo {author} {\bibfnamefont {M.}~\bibnamefont {Irfan}},\ and\
  \bibinfo {author} {\bibfnamefont {S.}~\bibnamefont {Qamar}},\ }\bibfield
  {title} {\bibinfo {title} {Multiqubit quantum phase gate using four-level
  superconducting quantum interference devices coupled to superconducting
  resonator},\ }\href@noop {} {\bibfield  {journal} {\bibinfo  {journal}
  {Physica C: Superconductivity}\ }\textbf {\bibinfo {volume} {477}},\ \bibinfo
  {pages} {24} (\bibinfo {year} {2012})}\BibitemShut {NoStop}%
\bibitem [{\citenamefont {Li}\ \emph {et~al.}(2015)\citenamefont {Li},
  \citenamefont {Ge}, \citenamefont {Liao},\ and\ \citenamefont
  {Feng}}]{EIT_4-level_SQC}%
  \BibitemOpen
  \bibfield  {author} {\bibinfo {author} {\bibfnamefont {H.}~\bibnamefont
  {Li}}, \bibinfo {author} {\bibfnamefont {G.}~\bibnamefont {Ge}}, \bibinfo
  {author} {\bibfnamefont {L.}~\bibnamefont {Liao}},\ and\ \bibinfo {author}
  {\bibfnamefont {S.}~\bibnamefont {Feng}},\ }\bibfield  {title} {\bibinfo
  {title} {Electromagnetically induced transparency and autler--townes
  splitting in a superconducting quantum circuit with a four-level v-type
  energy spectrum},\ }\href@noop {} {\bibfield  {journal} {\bibinfo  {journal}
  {Foundations of Physics}\ }\textbf {\bibinfo {volume} {45}},\ \bibinfo
  {pages} {198} (\bibinfo {year} {2015})}\BibitemShut {NoStop}%
\bibitem [{\citenamefont {Malik}\ \emph {et~al.}(2016)\citenamefont {Malik},
  \citenamefont {Erhard}, \citenamefont {Huber}, \citenamefont {Krenn},
  \citenamefont {Fickler},\ and\ \citenamefont {Zeilinger}}]{malik2016multi}%
  \BibitemOpen
  \bibfield  {author} {\bibinfo {author} {\bibfnamefont {M.}~\bibnamefont
  {Malik}}, \bibinfo {author} {\bibfnamefont {M.}~\bibnamefont {Erhard}},
  \bibinfo {author} {\bibfnamefont {M.}~\bibnamefont {Huber}}, \bibinfo
  {author} {\bibfnamefont {M.}~\bibnamefont {Krenn}}, \bibinfo {author}
  {\bibfnamefont {R.}~\bibnamefont {Fickler}},\ and\ \bibinfo {author}
  {\bibfnamefont {A.}~\bibnamefont {Zeilinger}},\ }\bibfield  {title} {\bibinfo
  {title} {Multi-photon entanglement in high dimensions},\ }\href@noop {}
  {\bibfield  {journal} {\bibinfo  {journal} {Nature Photonics}\ }\textbf
  {\bibinfo {volume} {10}},\ \bibinfo {pages} {248} (\bibinfo {year}
  {2016})}\BibitemShut {NoStop}%
\bibitem [{\citenamefont {Hu}\ \emph {et~al.}(2020)\citenamefont {Hu},
  \citenamefont {Xing}, \citenamefont {Zhang}, \citenamefont {Liu},
  \citenamefont {Pivoluska}, \citenamefont {Huber}, \citenamefont {Huang},
  \citenamefont {Li},\ and\ \citenamefont {Guo}}]{hu2020442}%
  \BibitemOpen
  \bibfield  {author} {\bibinfo {author} {\bibfnamefont {X.-M.}\ \bibnamefont
  {Hu}}, \bibinfo {author} {\bibfnamefont {W.-B.}\ \bibnamefont {Xing}},
  \bibinfo {author} {\bibfnamefont {C.}~\bibnamefont {Zhang}}, \bibinfo
  {author} {\bibfnamefont {B.-H.}\ \bibnamefont {Liu}}, \bibinfo {author}
  {\bibfnamefont {M.}~\bibnamefont {Pivoluska}}, \bibinfo {author}
  {\bibfnamefont {M.}~\bibnamefont {Huber}}, \bibinfo {author} {\bibfnamefont
  {Y.-F.}\ \bibnamefont {Huang}}, \bibinfo {author} {\bibfnamefont {C.-F.}\
  \bibnamefont {Li}},\ and\ \bibinfo {author} {\bibfnamefont {G.-C.}\
  \bibnamefont {Guo}},\ }\bibfield  {title} {\bibinfo {title} {Experimental
  creation of multi-photon high-dimensional layered quantum states},\
  }\href@noop {} {\bibfield  {journal} {\bibinfo  {journal} {npj Quantum
  Information}\ }\textbf {\bibinfo {volume} {6}},\ \bibinfo {pages} {1}
  (\bibinfo {year} {2020})}\BibitemShut {NoStop}%
\bibitem [{\citenamefont {Pucher}\ \emph {et~al.}(2022)\citenamefont {Pucher},
  \citenamefont {Liedl}, \citenamefont {Jin}, \citenamefont {Rauschenbeutel},\
  and\ \citenamefont {Schneeweiss}}]{raman_spin-controlled}%
  \BibitemOpen
  \bibfield  {author} {\bibinfo {author} {\bibfnamefont {S.}~\bibnamefont
  {Pucher}}, \bibinfo {author} {\bibfnamefont {C.}~\bibnamefont {Liedl}},
  \bibinfo {author} {\bibfnamefont {S.}~\bibnamefont {Jin}}, \bibinfo {author}
  {\bibfnamefont {A.}~\bibnamefont {Rauschenbeutel}},\ and\ \bibinfo {author}
  {\bibfnamefont {P.}~\bibnamefont {Schneeweiss}},\ }\bibfield  {title}
  {\bibinfo {title} {Atomic spin-controlled non-reciprocal raman amplification
  of fibre-guided light},\ }\href@noop {} {\bibfield  {journal} {\bibinfo
  {journal} {Nature Photonics}\ }\textbf {\bibinfo {volume} {16}},\ \bibinfo
  {pages} {380} (\bibinfo {year} {2022})}\BibitemShut {NoStop}%
\bibitem [{\citenamefont {Zheng}(1998)}]{multi-mode_cat-state}%
  \BibitemOpen
  \bibfield  {author} {\bibinfo {author} {\bibfnamefont {S.-B.}\ \bibnamefont
  {Zheng}},\ }\bibfield  {title} {\bibinfo {title} {A scheme for the generation
  of multi-mode schr{\"o}dinger cat states},\ }\href@noop {} {\bibfield
  {journal} {\bibinfo  {journal} {Quantum and Semiclassical Optics: Journal of
  the European Optical Society Part B}\ }\textbf {\bibinfo {volume} {10}},\
  \bibinfo {pages} {691} (\bibinfo {year} {1998})}\BibitemShut {NoStop}%
\bibitem [{\citenamefont {Enaki}\ \emph {et~al.}(2008)\citenamefont {Enaki},
  \citenamefont {Turcan},\ and\ \citenamefont {Vaseashta}}]{two-photon_laser}%
  \BibitemOpen
  \bibfield  {author} {\bibinfo {author} {\bibfnamefont {N.}~\bibnamefont
  {Enaki}}, \bibinfo {author} {\bibfnamefont {M.}~\bibnamefont {Turcan}},\ and\
  \bibinfo {author} {\bibfnamefont {A.}~\bibnamefont {Vaseashta}},\ }\bibfield
  {title} {\bibinfo {title} {Two photon multi mode laser model based on
  experimental observations},\ }\href@noop {} {\bibfield  {journal} {\bibinfo
  {journal} {J Optoelectron Adv Mater}\ }\textbf {\bibinfo {volume} {10}},\
  \bibinfo {pages} {3016} (\bibinfo {year} {2008})}\BibitemShut {NoStop}%
\bibitem [{\citenamefont {Stannigel}\ \emph {et~al.}(2012)\citenamefont
  {Stannigel}, \citenamefont {Rabl},\ and\ \citenamefont
  {Zoller}}]{driven-dissipative_preparation_entanglement}%
  \BibitemOpen
  \bibfield  {author} {\bibinfo {author} {\bibfnamefont {K.}~\bibnamefont
  {Stannigel}}, \bibinfo {author} {\bibfnamefont {P.}~\bibnamefont {Rabl}},\
  and\ \bibinfo {author} {\bibfnamefont {P.}~\bibnamefont {Zoller}},\
  }\bibfield  {title} {\bibinfo {title} {Driven-dissipative preparation of
  entangled states in cascaded quantum-optical networks},\ }\href@noop {}
  {\bibfield  {journal} {\bibinfo  {journal} {New Journal of Physics}\ }\textbf
  {\bibinfo {volume} {14}},\ \bibinfo {pages} {063014} (\bibinfo {year}
  {2012})}\BibitemShut {NoStop}%
\bibitem [{\citenamefont {Pichler}\ \emph {et~al.}(2015)\citenamefont
  {Pichler}, \citenamefont {Ramos}, \citenamefont {Daley},\ and\ \citenamefont
  {Zoller}}]{spin_networks}%
  \BibitemOpen
  \bibfield  {author} {\bibinfo {author} {\bibfnamefont {H.}~\bibnamefont
  {Pichler}}, \bibinfo {author} {\bibfnamefont {T.}~\bibnamefont {Ramos}},
  \bibinfo {author} {\bibfnamefont {A.~J.}\ \bibnamefont {Daley}},\ and\
  \bibinfo {author} {\bibfnamefont {P.}~\bibnamefont {Zoller}},\ }\bibfield
  {title} {\bibinfo {title} {Quantum optics of chiral spin networks},\
  }\href@noop {} {\bibfield  {journal} {\bibinfo  {journal} {Physical Review
  A}\ }\textbf {\bibinfo {volume} {91}},\ \bibinfo {pages} {042116} (\bibinfo
  {year} {2015})}\BibitemShut {NoStop}%
\end{thebibliography}%

\end{document}